\documentclass[a4paper, 12pt]{article}

\usepackage[latin1]{inputenc}
\usepackage[T1]{fontenc}
\usepackage{pslatex}
\usepackage{float}
\usepackage[intlimits,fleqn]{amsmath}
\usepackage{amsfonts}
\usepackage{amsbsy}
\usepackage{amssymb,dcolumn,exscale}
\usepackage{booktabs}
\usepackage{varioref}
\usepackage[width=16.5cm, left=2.2cm,top=2.5cm,height=24cm]{geometry}
\usepackage{cite}
\usepackage{graphicx,color}
\usepackage[hang]{subfigure}

\graphicspath{{Pics/}}
\DeclareGraphicsExtensions{.eps,.ps}

\newcommand{\as}{\alpha_s}
\newcommand{\yuk}{\hat{\alpha}_s}
\newcommand{\half}{\ensuremath{\tfrac{1}{2}}}
\newcommand{\mq}{m_{\tilde{q}}}
\newcommand{\mqL}{m_{\tilde{q}_L}}
\newcommand{\mqR}{m_{\tilde{q}_R}}
\newcommand{\mQL}{m_{\tilde{Q}_L}}
\newcommand{\mQR}{m_{\tilde{Q}_R}}
\newcommand{\mg}{m_{\tilde{g}}}
\newcommand{\sq}{\tilde{q}}
\newcommand{\sqb}{\tilde{q}^{\ast}}
\newcommand{\sQ}{\tilde{Q}}
\newcommand{\sQb}{\tilde{Q}^{\ast}}

\newcommand{\GeV}{\ensuremath{\,\mathrm{GeV}}}
\newcommand{\TeV}{\ensuremath{\,\mathrm{TeV}}}
\newcommand{\fb}{\ensuremath{\,\mathrm{fb}}}
\newcommand{\pb}{\ensuremath{\,\mathrm{pb}}}

\newcommand{\MSbar}{\ensuremath{\overline{\mbox{MS}}}}
\newcommand{\barq}{\ensuremath{\bar{q}}}
\newcommand{\hats}{\hat{s}}
\def\shat{{\hat s}}
\def\muf{{\mu^{}_f}}
\def\mufs{{\mu^{\,2}_f}}
\def\mur{{\mu^{}_r}}
\def\murs{{\mu^{\,2}_r}}
\newcommand{\gsim}{\raisebox{-0.07cm}{$\:\:\stackrel{>}{{\scriptstyle
 \sim}}\:\: $} }

\allowdisplaybreaks[2]

\begin{document}
%\maketitle
\begin{titlepage}
\noindent
DESY 09-004\\
SFB/CPP-09-01\\
Jan 2009 \hfill \today\\
\vspace{1.3cm}
\begin{center}
\Large{\bf 
Higher-order soft corrections to squark hadro-production
}\\
\vspace{1.5cm}
\large
U.~Langenfeld and S.~Moch\\[10mm]
\normalsize
{\it 
Deutsches Elektronensynchrotron DESY \\
\vspace{0.1cm}
Platanenallee 6, D--15738 Zeuthen, Germany}\\
\vspace{2.3cm}

\large
{\bf Abstract}
\vspace{-0.2cm}
\end{center}
We present new predictions for the total cross section of squark pair-production at Tevatron and LHC 
through next-to-next-to-leading order within the Minimal Supersymmetric Standard Model.
The results are based on the numerically dominant soft corrections. 
They are exact in all logarithmically enhanced terms near threshold, include
the Coulomb corrections at two loops and exact scale dependence.
We translate the increased total cross section at next-to-next-to-leading order 
into improved exclusion limits for squark masses 
and we investigate the scale dependence as well as the sensitivity on the parton luminosity.

\vfill
\end{titlepage}

\newpage

%
% ---------------------------------------------------------------------------
%
\section{Introduction}

Supersymmetry offers an attractive way for possible extensions of the Standard Model.
Its most popular incarnation for phenomenological studies 
is the Minimal Supersymmetric Standard Model (MSSM) which features 
a rather rich spectrum of new (heavy) particles. 
Active searches are currently being performed at the Tevatron and soon at LHC.
Squarks and gluinos as scalar and spin-$1/2$ supersymmetric partners 
of quarks and gluons could be pair-produced at a hadron collider at sizable rates 
(assuming $R$-parity).
So far, the Tevatron (CDF and D0 
collaboration~\cite{Peters:2008fn,:2007ww,Adams:2008nu}) 
has been providing lower limits on their masses depending
on certain assumptions about the parameter space of the MSSM.
At LHC in contrast, one expects for an initial luminosity of about $10\fb^{-1}$ 
and typical values of the MSSM parameters $10^6$ ($2500$) squark pairs 
for a mass of $300~\GeV$ ($1~\TeV$)~\cite{atlas:1999tdr2,Ball:2007zza}.
Given these large rates and the dedicated searches performed, it is natural 
to investigate the accuracy of the available theoretical predictions including 
quantum corrections.
\medskip

With squarks (and gluinos) carrying color charge, it is not surprising that 
Quantum Chromodynamics (QCD) provides the dominant corrections to the
production cross section.
This fact was realized some time ago and led to the computation 
of the complete next-to-leading order (NLO) QCD corrections~\cite{Beenakker:1996ch}.
The upshot is a large increase of the rate in comparison to 
leading order (LO) QCD predictions 
along with a reduced scale dependence indicating the improved theoretical uncertainty.
The origin of these large higher order QCD corrections in hadro-production of
(heavy) colored particles is well known, since it is related to universal QCD dynamics. 
As a typical pattern so-called Sudakov logarithms show up which originate 
from soft gluon emission in regions of phase space near the 
partonic threshold. 
They depend on the squark velocity $\beta=(1-4\mq^2/\hats)^{1/2}$ and become large for 
center-of-mass energies ${\sqrt{\hats}}$ near the threshold for 
squark pair-production, $\hats \simeq 4\mq^2$. 
Sudakov logarithms can be organized to all orders of perturbation theory by means 
of a threshold resummed cross section to a given logarithmic accuracy.
\medskip

In this letter, we study soft gluon effects for the total cross section of squark hadro-production.
We employ Sudakov resummation to generate approximate next-to-next-to-leading order (NNLO) 
QCD predictions which are accurate in all $\log(\beta)$-enhanced terms at two loops. 
Moreover, we include the complete two-loop Coulomb 
corrections as well as the exact dependence on the renormalization and factorization scale.
To that end, we largely follow a similar study for top-quark hadro-production at the Tevatron
and the LHC~\cite{Moch:2008qy,Moch:2008ai}. 
The importance of Sudakov resummation for the latter reaction has often been
emphasized in the literature (see e.g. Ref.~\cite{Laenen:2008jx}).
Squarks are generally believed to be heavier than top quarks, but light enough
to be pair-produced at these colliders as well, 
so that the need for threshold resummation 
(and the technique) carries over from top-quark pair-production.
Recently, the soft gluon resummation for squark and gluino hadro-production 
has been performed to next-to-leading logarithmic (NLL) accuracy in Ref.~\cite{Kulesza:2008jb},
and results compatible with ours were found.
See also Refs.~\cite{Hollik:2008yi,Hollik:2007wf,Bornhauser:2007bf} for the
electroweak contributions to (top)-squark pair-production through NLO.
\medskip

The letter is organized as follows.
We recall the dominant parton channels contributing to the cross section of squark pair-production. 
Then we discuss the NLO QCD corrections. 
For the latter, we have determined fits to the exact NLO scaling functions 
for representative choices of squark and gluino masses (presented in the Appendix).
Next, we describe the steps necessary to achieve soft gluon resummation 
to next-to-next-to-leading logarithmic (NNLL) accuracy 
where, we focus on the differences with respect to the well-known procedure for top-quarks. 
This extends the NLL results of Ref.~\cite{Kulesza:2008jb}.
We employ the resummed cross section to derive new (approximate) NNLO expressions 
for the scaling functions and analyze the dependence of the hadronic cross section on the scale 
and on parton distribution functions (PDFs).
For the finite order expansion to NNLO, we find good apparent convergence
properties and a markedly improved stability of the total cross section with
respect to scale variations.
We give two examples, how our results translate into new limits on the squark masses.

%
% ---------------------------------------------------------------------------
%
\section{Setting the stage}

\begin{figure}[ht]{
  \centering
  {
    \scalebox{1.0}{\includegraphics[clip]{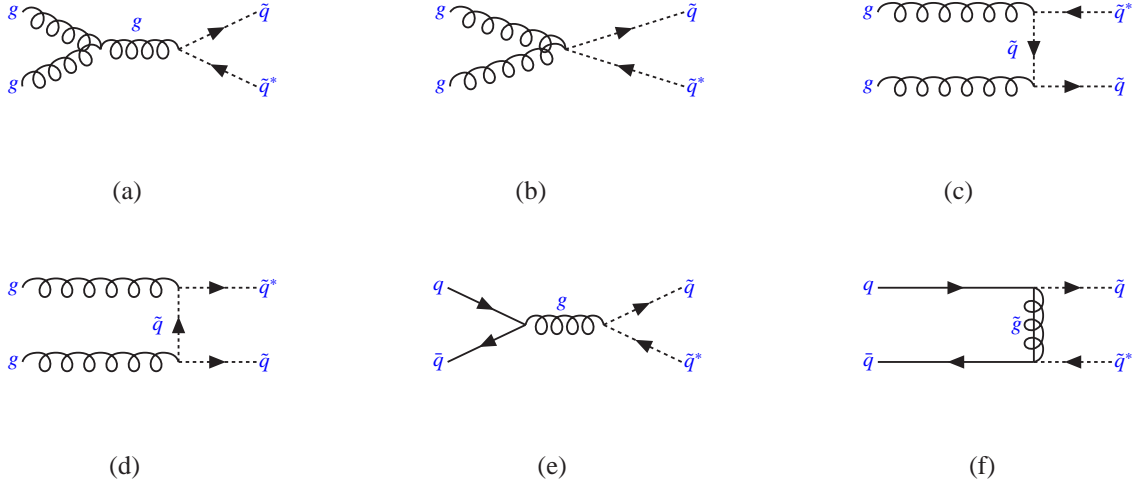}} 
  }
  \caption{\small
  \label{fig:ggtosqsqbar}
  Contributing diagrams to the processes $gg \to \sq\sqb$
  (diagrams~(a)--(d)) 
  and $q\bar{q} \to \sq\sqb$
  (diagrams~(e),(f)).
  Diagram (f) is proportional to the squark-gluino Yukawa coupling $\yuk$ 
  and introduces dependence on the gluino mass $\mg$.}
}
\end{figure}

We focus on the inclusive hadronic cross section of hadro-production of squark pairs, 
$\sigma_{p p \rightarrow \sq \sqb X}$ which is a function 
of the hadronic center-of-mass energy $\sqrt{s}$, the squark mass $\mq$, 
and the gluino mass $\mg$.
In the standard factorization approach of perturbative QCD, it reads
\begin{eqnarray}
  \label{eq:totalcrs}
  \sigma_{pp \to \sq \sqb X}(s,\mq^2,\mg^2) &=& 
  \sum\limits_{i,j = q,{\bar{q}},g} \,\,\,
  \int\limits_{4\mq^2}^{s }\,
  d {\shat} \,\, L_{ij}(\shat, s, \mufs)\,\,
  \hat{\sigma}_{ij \to \sq \sqb} ({\shat},\mq^2,\mg^2,\mufs,\murs)\, ,
\end{eqnarray}
where the parton luminosities $L_{ij}$ are given as convolutions 
of the PDFs $f_{i/p}$ defined through
\begin{eqnarray}
  \label{eq:partonlumi}
  L_{ij}(\shat, s, \mufs) &=& 
  {\frac{1}{s}} \int\limits_{\shat}^s 
  {\frac{dz}{z}} f_{i/p}\left(\mufs,{\frac{z}{s}}\right) 
  f_{j/p}\left(\mufs,{\frac{\shat}{z}}\right)\, ,
\end{eqnarray}
where $\shat$ denotes the partonic center of mass energy. 
Factorization and renormalization scales $\muf$ and $\mur$ are identified 
($\muf = \mur \equiv \mu$).
The sum in Eq.~(\ref{eq:totalcrs}) runs over 
all massless parton flavors ($n_f=5$ for LHC and Tevatron).
The hard parton scattering cross section $\hat{\sigma}_{ij \to \sq \sqb}$ 
appearing in Eq.~(\ref{eq:totalcrs}) receives at Born level contributions from 
the channels
\begin{eqnarray}
  \label{eq:Borngg}
  g\, g 
  &\rightarrow& \sq_k\,\sqb_l
\, ,
\\
  \label{eq:Bornqq}
  q_i\, \bar{q}_j 
  &\rightarrow& \sq_k\,\sqb_l
\, ,
\end{eqnarray}
where $i$, $j$, $k$ and $l$ denote flavor indices. 
For the calculation of the hadronic cross section~(\ref{eq:totalcrs}) in this paper, 
we always sum over all final state squark flavors allowed by quantum number
conservation in proton-(anti)-proton scattering except for top-squarks (stops).
The contributing diagrams are displayed in Fig.~\ref{fig:ggtosqsqbar}.

While the gluon fusion process~(\ref{eq:Borngg}) always leads 
to like flavor squark pairs ($k=l$) and identical chiralities 
(see e.g. Ref.~\cite{Beenakker:1996ch} for details on the quantum numbers),
a $q \bar{q}$ initial state leads to a more interesting flavor structure 
of the final state squark pair.
For the $q \bar{q}$-scattering process~(\ref{eq:Bornqq}) 
we have final states with the same flavor structure 
if the flavors of the initial state are different ($i \neq j \Rightarrow i=k \wedge j=l$) 
due to a gluino exchange in the $t$-channel.
Likewise, if the initial flavors are equal ($i=j$), then the final flavors 
may be different from the initial flavors,  
but the flavors of the $\sq \sqb$-pair are equal due to a gluon
exchange in the $s$-channel ($i=j \Rightarrow k = l$).  
Moreover, if we keep the squark chirality as a second quantum number, we 
encounter a richer structure and much of the discussion 
(including soft corrections) carries over. 
For the subsequent study of NLO and approximate NNLO perturbative QCD, 
we restrict ourselves to cross sections summed over all possible final squark flavors 
and chiralities ($L,R$).
The complete list of Born cross sections for different flavors and chiralities 
for the process~(\ref{eq:Bornqq}) is presented in the Appendix.
In the following, we focus on the total partonic Born cross sections $\hat{\sigma}^B$
where all flavors and chiralities are summed over, using the conditions $i=j$ or $i\neq j$
(see e.g. Ref.~\cite{Beenakker:1996ch}),
\begin{eqnarray}
  \label{eq:totalgg}
  \hat{\sigma}^B_{gg \to \sq \sqb}(\hats,\mq^2) &\!=\!\!&
  \frac{\alpha_s^2}{\mq^2}\*{\frac {n_f}{192}}\,\pi \, \left( 1-{\beta}^{2} \right)\!\!
  \left[ 41\,\beta-31\,{\beta}^{3} + 
    \left(17 - 18 \beta^2 + \beta^4 \right)\,
    \log \left({\frac {1-\beta}{1+\beta}} \right)  \right]
  \, \!,
  \qquad
  \\[2ex]
  \label{eq:totalqq}
  \hat{\sigma}^B_{q_i \bar{q}_j \to \sq \sqb}(\hats,\mq^2,\mg^2) &\!=\!\!&
  \delta_{ij}n_f \frac{4\pi\as^2}{27 \hats}\beta^3 
  \nonumber\\[3mm]
  & & 
  -\frac{4\pi\yuk^2}{9 \hats}
  \left[\beta \left(1 + \frac{(\mg^2 - \mq^2)^2}
      {\mg^2 \hats + (\mg^2 - \mq^2)^2}\right)
    + \left(1 + 2 \frac{\mg^2 - \mq^2}{\hats}\right)L_1\right] 
  \nonumber\\[3mm]
  & & 
  + \delta_{ij} \frac{4\pi\as\yuk}{27 \hats}
  \left[\beta\left(1 + 2 \frac{\mg^2 - \mq^2}{\hats}\right)
    +2 \left(\frac{\mg^2}{\hats} + \frac{(\mg^2 - \mq^2)^2}{\hats^2}\right)L_1
  \right]
  \, ,
\end{eqnarray}
with the strong coupling constant $\as$ and the squark-gluino Yukawa coupling
$\yuk$ identical to $\as$ as required by supersymmetry (see the discussion below). 
Moreover, we have abbreviated 
\begin{eqnarray}
  \label{eq:L1def}
  L_1 &=& \log\left(\frac{\hats + 2(\mg^2 - \mq^2) - \hats\beta}
                         {\hats + 2(\mg^2 - \mq^2) + \hats\beta}\right)
  \, ,
\end{eqnarray}
where $\beta=(1-4\mq^2/\hats)^{1/2}$ is the squark velocity. 

For the studies of higher order QCD corrections, 
the partonic cross section can be expressed in terms of scaling functions 
$f_{ij}$. For gluon fusion, we define 
\begin{eqnarray}
  \label{eq:ggkl}
  \hat{\sigma}_{gg \to \sq \sqb} (\hats,\mq^2) 
  &=& \frac{\alpha_s^2}{\mq^2}\, 
  \sum\limits_{k=0}^\infty (4 \pi \alpha_s)^k\, 
  \sum\limits_{l=0}^k\, 
  f_{gg}^{(kl)}(\eta)\, 
  \log^l\left(\frac{\mu^2}{\mq^2}\right)
  \, ,
\end{eqnarray}
and, likewise, for $q \bar{q}$-scattering $f_{q_i \bar{q}_j}^{(kl)}$ with
additional dependence on the gluino mass $\mg$, see Eq.~(\ref{eq:totalqq}).
Here $\eta = \hats/(4 \mq^2) -1$ is a measure for the distance
to the production threshold at $\hats = 4\mq^2$.
The NLO QCD corrections are known~\cite{Beenakker:1996ch} although they are
not available in analytical form (unlike the case of 
top-quark hadro-production~\cite{Czakon:2008ii}).
Instead in Ref.~\cite{Beenakker:1996ed}, the authors present a numerical program 
called {\sc{Prospino}} to calculate the complete hadronic cross section at NLO. 
We have used this program to extract values for the scaling functions 
$f^{(10)}_{gg}$, $f^{(10)}_{q\barq}$, and $f^{(10)}_{q_i \bar{q}_j}$ 
(all terms proportional to $\log(\mu^2/\mq^2)$ will be discussed below). 
Subsequently, we have determined fits of these functions to per mille accuracy 
based on the following ansatz,
\begin{eqnarray}
  \label{eq:fit10}
  f_{gg}^{(10)} &=& 
  \frac{7 n_f}{192\pi} \beta\,
  \left( \frac{3}{2} \log^2 \left( 8\,{\beta}^{2} \right)
    - \frac{183}{28} \log \left( 8\,{\beta}^{2} \right) 
    + \frac{11\*\pi^2}{336\*\beta}\right) + h_{gg}(\beta)
  \, ,\\[2ex]
  \label{eq:fitqq10}
  f_{q_{i}\bar{q}_{j}}^{(10)} &=&
  \frac{4}{9\pi}\frac{m_{\tilde{g}}^2 m_{\tilde{q}}^2}
                                    {(m_{\tilde{g}}^2 + m_{\tilde{q}}^2)^2}
                                    \beta\, 
  \left(\frac{2}{3} \log^2(8\beta^2) 
    - \frac{11}{4} \log(8\beta^2) 
    + \frac{7\pi^2}{48\beta}\right) + h_{q_{i}\bar{q}_{j}}(\beta)
  \, .
\end{eqnarray}
The threshold logarithms $\log(\beta)$ as well as 
the Coulomb corrections ($\sim 1/\beta$) are kept
exactly~\cite{Beenakker:1996ch}, while the 
fit function $h(\beta)$ is given in Eq.~(\ref{eq:hfitdef}). 
Near threshold, $h(\beta) = a_1 \beta + {\cal O}(\beta^2)$ 
holds (see Tab.~\ref{tab:fitcoeff}) and 
these fits allow for an easy handling of the NLO QCD corrections in
phenomenological applications.

The $\log (\beta)$ terms appearing in Eqs.~(\ref{eq:fit10}) and (\ref{eq:fitqq10}) 
can be resummed systematically to all orders in perturbation theory 
employing the well established techniques (see e.g.
Refs.~\cite{Contopanagos:1997nh,Catani:1996yz,Moch:2005ba}).
The resummation typically proceeds in Mellin space after introducing moments $N$ 
with respect to the variable $\rho = 4\mq^2 /\hats$ as
\begin{eqnarray}
  \label{eq:mellindef}
  \hat{\sigma}(N,\mq^2,\mg^2) &=&
  \int\limits_{0}^{1}\,d\rho\, \rho^{N-1}\,
  \hat{\sigma}(\hats,\mq^2,\mg^2)\, .
\end{eqnarray}
For scattering reactions with non-trivial color exchange,
one  has to choose a suitable color basis for the total cross-section.
It is convenient to select a decomposition according to color-singlet
and color-octet final states,
\begin{eqnarray}
  \label{eq:singlet-octet}
  \hat{\sigma}_{ij \to \sq \sqb} (\hats,\mq^2,\mg^2) 
  &=& 
  \sum\limits_{\bf I=1,8}\, \hat{\sigma}_{ij,\, {\bf I}}(\hats,\mq^2,\mg^2)\, .
\end{eqnarray}
At Born level, we find the singlet component to be explicitly given by
\begin{eqnarray}
  \label{eq:Bornsinglet}
{\hat{\sigma}}_{gg,\, {\bf 1}}^B &=&
       {\frac{\as^2}{\mq^2}}\* {\frac{n_f\, \pi \rho}{4 N_c (N_c^2 -1)}}\*
       \left[- \beta^3 + 2 \beta + {\frac{1}{2}}\left(1 - \beta^4\right)
       \*\log\left({\frac{1-\beta}{1+\beta}}\right)\right]
     \, ,
     \\[2ex]
{\hat{\sigma}}_{q_i\bar{q}_j,\, {\bf 1}}^B &=&
      - \frac{\yuk^2}{\mq^2}\frac{C_F^2\pi}{2 N_c^2}\rho
      \left[\beta \left(1 + \frac{(\mg^2 - \mq^2)^2}{\mg^2 \hats + (\mg^2 - \mq^2)^2}\right)
      + \left(1 + 2 \frac{\mg^2 - \mq^2}{\hats}\right)L_1\right]
    \, ,
\end{eqnarray}
and the octet terms ${\hat{\sigma}}_{ij,\, {\bf 8}}$ 
can be easily derived from Eqs.~(\ref{eq:totalgg}) and (\ref{eq:totalqq}).
The resummed cross sections (defined in the \MSbar-scheme), 
for the individual color structures of the scattering process 
are then obtained as single exponentials in Mellin-space, 
\begin{eqnarray}
  \label{eq:sigmaNres}
  \frac{\hat{\sigma}_{ij,\, {\bf I}}(N,\mq^2,\mg^2) }
       { \hat{\sigma}^{B}_{ij,\, {\bf I}}(N,\mq^2,\mg^2)} 
  &=& 
  g^0_{ij,\, {\bf I}}(\mq^2,\mg^2) \cdot \exp\, \left( G_{ij,\,{\bf I}}(N+1) \right) + 
  {\cal O}(N^{-1}\log^n N) \, ,
\end{eqnarray}
where all dependence on the renormalization and
factorization scale $\mur$ and $\muf$ is suppressed and  
the respective Born term is denoted $\hat{\sigma}^{B}_{ij,\, {\bf I}}$.
The exponent $G_{ij,\, {\bf I}}$ contains all large Sudakov logarithms $\log^k N$ 
and the resummed cross section~(\ref{eq:sigmaNres}) is 
accurate up to terms which vanish as a power for large Mellin-$N$.
To NNLL accuracy, $G_{ij,\, {\bf I}}$ is commonly written as 
\begin{eqnarray}
  \label{eq:GNexp}
  G_{ij,\, {\bf I}}(N) = 
  \log N \cdot g^1_{ij}(\lambda)  +  g^2_{ij,\, {\bf I}}(\lambda)  + 
  \frac{\alpha_s }{ 4 \pi}\, g^3_{ij,\, {\bf I}}(\lambda)  + \dots\, ,
\end{eqnarray}
where $\lambda = \beta_0\, \log N\, \alpha_s/(4 \pi)$. 
The functions $g^k_{ij}$ for the singlet and octet color structures are explicitly given 
in Ref.~\cite{Moch:2008qy} and can be taken over from the case of top-quark
hadro-production (see also Ref.~\cite{Kulesza:2008jb} for the results to NLL accuracy).
All $g^k_{ij}$, $k=0,...,3$ depend on a number of 
anomalous dimensions, i.e. the well-known cusp anomalous dimension $A$,
the functions $D$ and $D_{q{\bar q}}$ controlling soft emission,
and the coefficients of the QCD $\beta$-function.
At higher orders their precise expressions also depend on the chosen renormalization scheme,  
thus on the dynamical degrees of freedom.
At the center-of-mass energies of Tevatron and LHC and for the 
mass ranges currently considered in MSSM phenomenology 
$\mq, \mg \simeq {\cal O}(200\GeV - 1\TeV)$, 
a scheme is appropriate which decouples all heavy particles (top-quark, squarks, gluino).
Thus, we are left with the Standard Model $\beta$-function coefficients 
$\beta_0 = 11 - (2/3) n_f$ and $\beta_1 = 102 - (38/3) n_f$ 
and the same expressions 
for the anomalous dimension $A$, $D$ and $D_{q{\bar q}}$ as in the case of
top-quark hadro-production (see  Ref.~\cite{Moch:2008qy}).
Our scheme choice is in line with the exact NLO QCD calculation~\cite{Beenakker:1996ch} 
to facilitate matching (see below).
Throughout the paper, $\mq$ and $\mg$ denote pole masses.

Having Eq.~(\ref{eq:sigmaNres}) and all quantities 
necessary for its explicit evaluation at our disposal,
we use the resummed cross section as a generating functional for 
the threshold approximation to the yet unknown NNLO QCD corrections,
i.e. the inclusive partonic scaling functions $f^{(20)}_{ij}$ in Eq.~(\ref{eq:ggkl}).
Substituting all numerical values and setting $n_f = 5$, 
we obtain in the \MSbar-scheme the approximate NNLO results,
\begin{eqnarray}
\label{eq:fgg20-num}
  f^{(20)}_{gg} &=& 
  \frac{f^{(00)}_{gg}}{(16\pi^2)^2}\*
  \Biggr\{ 
            4608 \* \log^4 \beta
          - 1894.9144 \* \log^3 \beta
\nonumber\\
          &&
       + \left(
          - 14306.950
          + 26471.239 \* a_{1, gg}
%         - 1309.5423
          + 496.30011 \* {\frac{1}{\beta}} \right) \* \log^2 \beta 
\nonumber\\
          &&
       + \left( 
            2491.3858
          + 2102.9161 \* a_{1, gg}
%           3523.9200
          + 321.13660 \* {\frac{1}{\beta}} \right) \* \log \beta
\nonumber\\
          &&
          + 68.547138 \* {\frac{1}{\beta^2}}
          - 196.93242 \* {\frac{1}{\beta}}
          + C^{(2)}_{gg}
    \Biggr\}
    \, ,\qquad
    \\
%\end{eqnarray}
%
%\begin{eqnarray}
\label{eq:fq1q220-num}
  f^{(20)}_{q_i\bar{q}_j} &=&
  \frac{f^{(00)}_{q_i\bar{q}_j}}{(16\pi^2)^2}\*
  \Biggl\{ 
           \frac{8192}{9} \* \log^4 \beta
          - 405.30701 \* \log^3 \beta
\nonumber\\
          &&
       + \left( 
          - 2474.3762
          + 4825.4863 \* \frac{(1+r_q^2)^2 }{ r_q^2} \* a_{1, q_i{\bar q}_j}(r_q)
%           450.41183
          + 982.57395 \* {\frac{1}{\beta}} \right) \* \log^2 \beta 
\nonumber\\
          &&
       + \left( 
            691.27647
          + 383.34407 \* \frac{(1+r_q^2)^2}{r_q^2} \* a_{1, q_i{\bar q}_j}(r_q)
%           923.62614
          + 336.72883 \* {\frac{1}{\beta}} \right) \* \log \beta
\nonumber\\
          &&
          + 205.64141 \* {\frac{1}{\beta^2}}
          - 634.79109 \* {\frac{1}{\beta}}
          + C^{(2)}_{q_i{\bar q}_j}\Biggr\} 
    \, ,\qquad
\end{eqnarray}
where the coefficients of the $\log^4 \beta$-terms are exact 
and the constants $a_1$ from the NLO matching are given in Tab.~\ref{tab:fitcoeff}, 
respectively Eq.~(\ref{eq:a1rqdef}).
In the derivation we have used the linearization of the Born functions, 
\begin{eqnarray}
  \label{eq:fgg00}
  f^{(00)}_{gg} &=& 
  {\frac {7 n_f}{192}}\,\pi \beta + \mathcal{O}(\beta^3) 
  \, ,
  \\
  \label{eq:fqq00}
  f^{(00)}_{q_i\bar{q}_j}&=& 
  \frac{4}{9}\, \pi \beta \frac{\mg^2 \mq^2}{(\mg^2 + \mq^2)^2} +
  \mathcal{O}(\beta^3)
  \, .
\end{eqnarray}
For our phenomenological studies though, we have always substituted the full Born
result for $f^{(00)}_{ij}$ in Eqs.~(\ref{eq:fgg20-num})--(\ref{eq:fq1q220-num}).

A few comments on Eqs.~(\ref{eq:fgg20-num})--(\ref{eq:fq1q220-num}) are in order here.
The results are accurate to all powers in $\log(\beta)$ at two loops.
This has been achieved by consistent matching of the resummed cross section to the 
exact NLO result. 
To that end, we have used our fits~(\ref{eq:fit10}) and (\ref{eq:fitqq10}),
specifically the constants $a_1$ of Eq.~(\ref{eq:hfitdef}) 
as given in Tab.~\ref{tab:fitcoeff}.
These constants enter in the coefficients of $\log^2(\beta)$  and $\log(\beta)$ in 
Eqs.~(\ref{eq:fgg20-num})--(\ref{eq:fq1q220-num}), i.e. the quadratic and
linear logarithm, 
so that the first two terms become $(- 1309.5423, 3523.9200)$ in Eq.~(\ref{eq:fgg20-num}).
Likewise for a representative choice of squark and gluino masses 
$\mq = 400\GeV$ and $\mg = 500\GeV$,
we find the first two terms of $\log^2(\beta)$  and $\log(\beta)$ to combine 
into $(567.40606, 932.92034)$,
if the quarks in the initial state have identical flavors, 
and into $(450.41183, 923.62614)$ in Eq.~(\ref{eq:fq1q220-num}),
if the quarks in the initial state have different flavors, 
see also Eqs.~(\ref{eq:fqq00}) and (\ref{eq:a1rqdef}).
Strictly speaking, for the NLO matching separate constants are required 
for the singlet and octet color structures at NLO 
and for this reason the exact numerical coefficient of the term linear in $\log \beta$ 
will differ slightly. 
However, experience from studies for top-quark hadro-production shows 
this effect to be marginal~\cite{Czakon:2008ii} 
and well covered within our quoted overall uncertainty for the threshold approximation.
On top of the threshold logarithms we have also added in
 Eqs.~(\ref{eq:fgg20-num})--(\ref{eq:fq1q220-num}) 
the complete two-loop Coulomb corrections (as summarized e.g. in Ref.~\cite{Moch:2008qy}).
Finally, we comment on the presence of the squark-gluino Yukawa coupling
$\yuk$, which is identified with the strong coupling constant $\as$ 
as required by supersymmetry. 
In the \MSbar-scheme this is achieved by a (finite) renormalization, which is
correctly implemented at one-loop level by our matching procedure at NLO.
At the two-loop level, the additional renormalization of $\yuk$ 
would only affect the constant terms { $C^{(2)}_{q\barq}$ and} 
$C^{(2)}_{q_i{\bar q}_j}$ in Eq.~(\ref{eq:fq1q220-num}).
Since we have no control over { these constants}, 
we set all two-loop constants $C^{(2)}_{ij}$ to zero in our phenomenological studies.

\bigskip

At this stage, it only remains to discuss those terms in Eq.~(\ref{eq:ggkl}) which
describe scale dependence, i.e. the terms proportional to $\log (\mu^2/\mq^2)$.
Through NNLO, it concerns $f^{(11)}_{q_i\barq_{j}}$, $f^{(21)}_{q_i\barq_{j}}$ 
and $f^{(22)}_{q_i\barq_{j}}$ 
which are entirely determined by renormalization group arguments
(see. e.g. Refs.~\cite{vanNeerven:2000uj,Kidonakis:2001nj}).
They can be constructed with the help of lower order results, that is the
scaling functions $f^{(00)}_{q_i\barq_{j}}$ and $f^{(10)}_{q_i\barq_{j}}$ 
and the splitting functions $P_{ij}$.
The latter quantities govern the PDF evolution.
They can be expanded as
\begin{eqnarray}
  \label{eq:Psplitting}
  P_{ij}(x) &=& \frac{\alpha_s }{ 4 \pi} \* P^{(0)}_{ij}(x) + 
  \left( \frac{\alpha_s }{ 4 \pi} \right)^2 \* P^{(1)}_{ij}(x) + \dots
  \, ,
\end{eqnarray}
and explicit expressions for the $P_{ij}^{(k)}$ can be found in Refs.~\cite{Moch:2004pa,Vogt:2004mw}.
Following Refs.~\cite{vanNeerven:2000uj,Kidonakis:2001nj}, 
we can calculate the scale dependence of the partonic cross section. 
Again, we have identified renormalization and factorization scale 
$\mu_r = \mu_f = \mu$. 
However, the factorization scale dependence can easily be recovered
by re-expansion of $\alpha_s(\mu_f)$ in terms of  $\alpha_s(\mu_r)$. 
This yields the following result
\begin{eqnarray}
\label{eq:f11}
f_{ij}^{(11)}&=&
\frac{1}{16\pi^2}\*\left(
  2\* \beta_0 \*f_{ij}^{(00)} - f_{kj}^{(00)}\otimes P_{ki}^{(0)} - f_{ik}^{(00)}\otimes P_{kj}^{(0)}
\right)
\, ,\\[2mm]
\label{eq:f21}
f_{ij}^{(21)}&=& 
\frac{1}{(16\pi^2)^2}\*\left(
  2\* \beta_1\* f_{ij}^{(00)} - f_{kj}^{(00)}\otimes P_{ki}^{(1)} 
  - f_{ik}^{(00)}\otimes P_{kj}^{(1)}\right)
\nonumber\\*
& &
+ \frac{1}{16\pi^2}\*\left(
  3 \*\beta_0 \*f_{ij}^{(10)} 
  -f_{kj}^{(10)}\otimes P_{ki}^{(0)}
  - f_{ik}^{(10)}\otimes
  P_{kj}^{(0)}\right)
\, ,
\\[2mm]
\label{eq:f22}
f_{ij}^{(22)}&=&
\frac{1}{(16\pi^2)^2}\*\left(
  f_{kl}^{(00)}\otimes P_{ki}^{(0)}\otimes P_{lj}^{(0)}
  +\frac{1}{2} f_{in}^{(00)}\otimes P_{nl}^{(0)}\otimes P_{lj}^{(0)}
  +\frac{1}{2} f_{nj}^{(00)}\otimes P_{nk}^{(0)}\otimes P_{ki}^{(0)}\right.\notag\\*[2mm]
& & \hspace{18mm}\left. + 3 \*\beta_0^2  \*f_{ij}^{(00)} 
  - \frac{5}{2}\*\beta_0 \*f_{ik}^{(00)}\otimes P_{kj}^{(0)}
  - \frac{5}{2}\*\beta_0 \*f_{kj}^{(00)}\otimes P_{ki}^{(0)}
\right)
\, ,			
\end{eqnarray}
where $\otimes$ denotes the standard Mellin convolution, i.e. ordinary
products in Mellin space under transformation~(\ref{eq:mellindef}).
The summation over repeated indices for admissible parton contributions 
is implied, although for phenomenological applications 
we restrict ourselves in Eqs.~(\ref{eq:f21}) and (\ref{eq:f22}) 
to the (numerically dominant) diagonal parton channels at two loops.
Note, that the scale dependent two-loop contributions are exact at all
energies also away from threshold.
In this context, we remark that the usual definition of the splitting function $P_{qg}$ 
(see e.g. Refs.~\cite{Moch:2004pa,Vogt:2004mw}) needs to be divided with a factor $2n_f$, 
because the standard definition implicitly contains a sum over all flavors.
One other comment concerns the two-loop splitting function $P_{qq}$ in Eq.~(\ref{eq:f21}).
Obviously, only the flavor non-singlet splitting function $P_{qq} = P^+_{ns}$
enters the convolution~(\ref{eq:f21}) for the flavor non-singlet scaling function
$f^{(21)}_{q_i\bar{q}_j}$ because flavor number is conserved.
Likewise for the flavor singlet case $f^{(21)}_{q\bar{q}}$,
the flavor singlet splitting function $P_{qq} = P^+_{ns} +  P_{ps}$ 
(i.e. the sum of non-singlet and pure-singlet) contributes.
Finally, $f^{(00)}_{q_i\bar{q}_j}$ depends only on the ratio of
the gluino and squark mass $r_q = \mg/\mq$.
Therefore through Eqs.~(\ref{eq:f11}) and (\ref{eq:f22}),
the scaling functions $f^{(11)}_{q_i\bar{q}_j}$ and $f^{(22)}_{q_i\bar{q}_j}$ 
depend only on this ratio.

%
% ---------------------------------------------------------------------------
%
\section{Numerical Results}

\begin{figure}
\centering
\subfigure[\label{subfig:fqq1}]
    {
        \scalebox{0.32}{\includegraphics{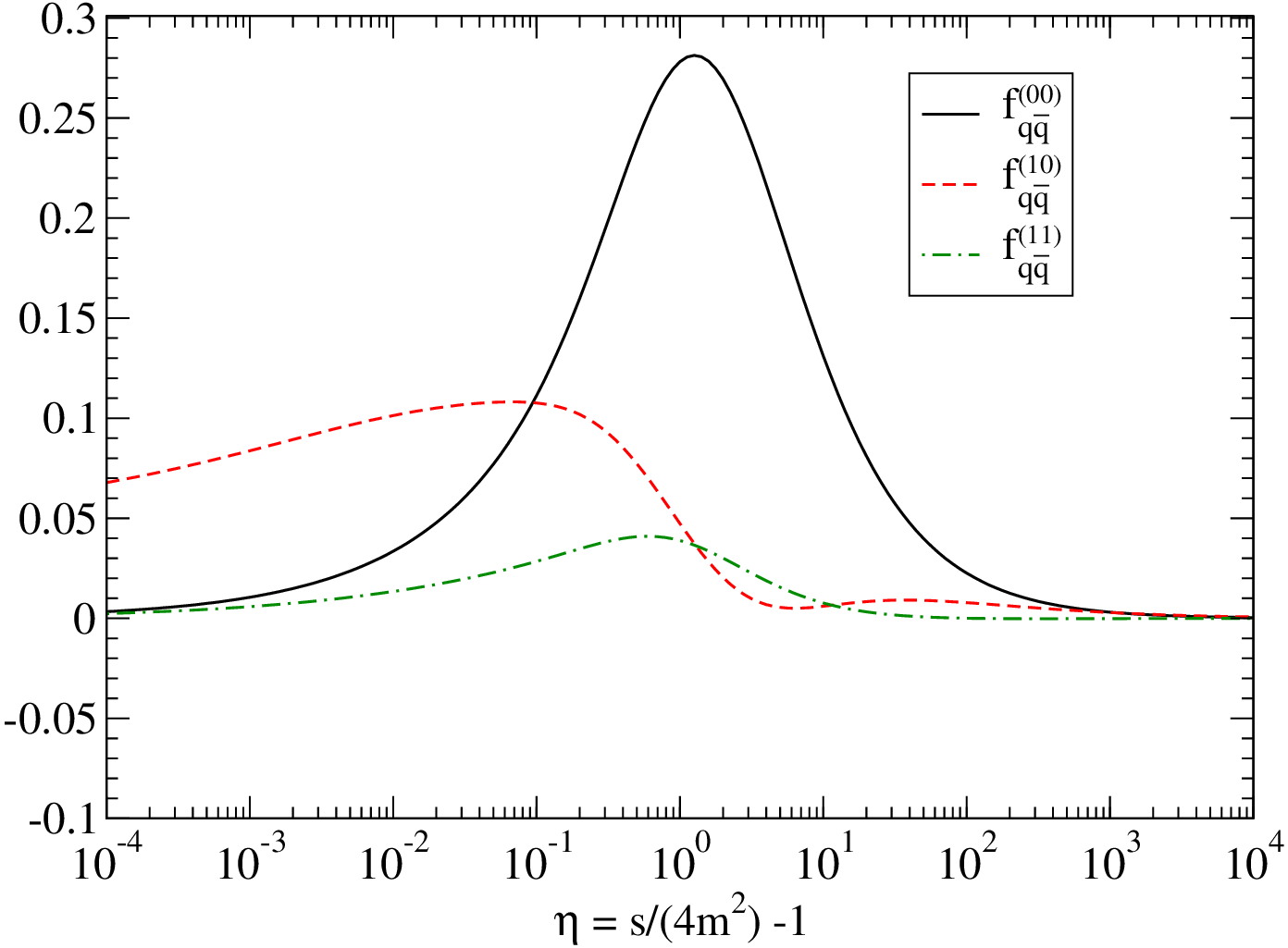}}
    }
\hspace*{3mm}
\subfigure[\label{subfig:fq1q21}]
    {
       \scalebox{0.32}{\includegraphics{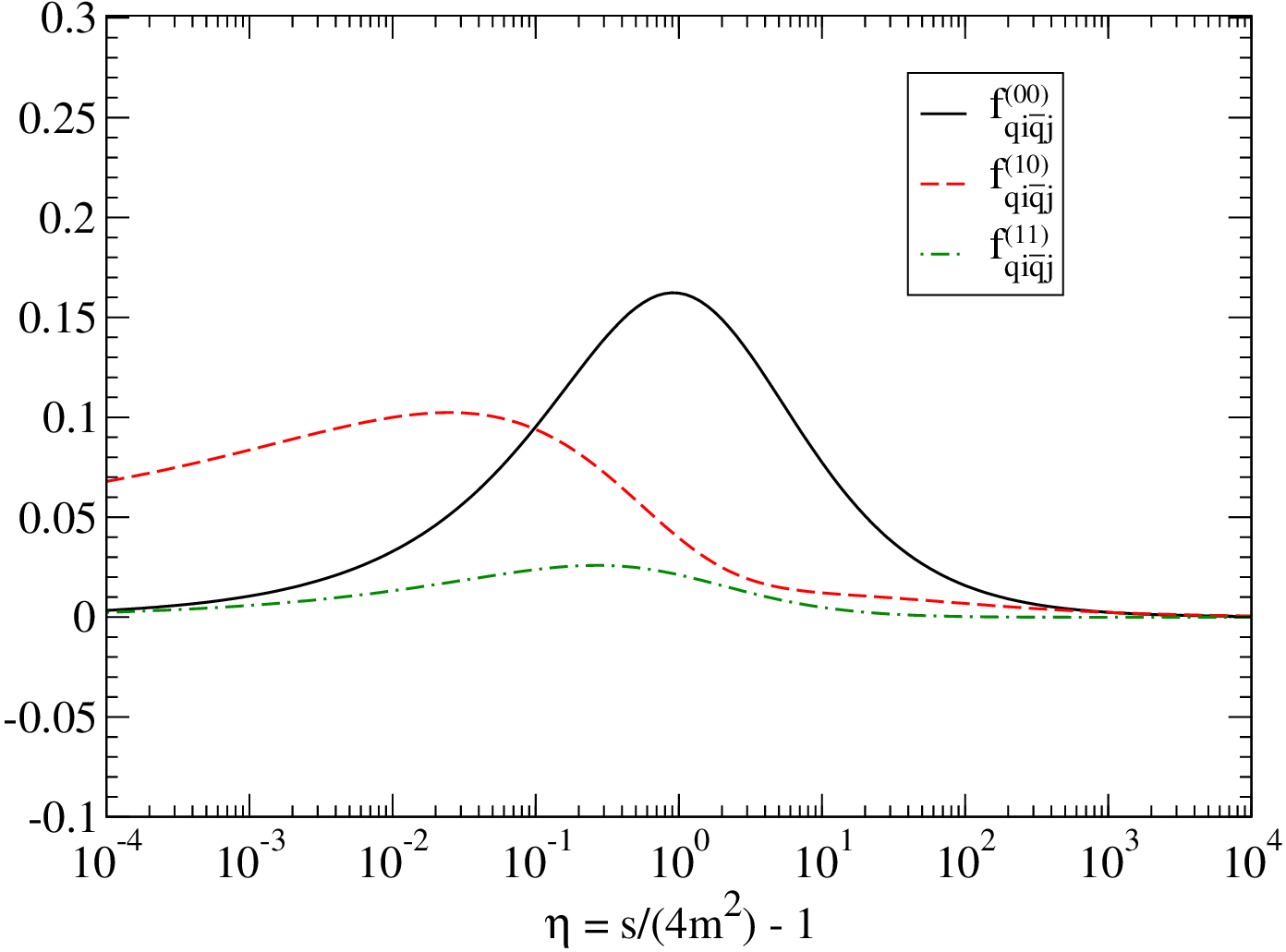}} 
    }
\vspace*{6mm}

\subfigure[\label{subfig:fqq2}]
    {
        \scalebox{0.32}{\includegraphics{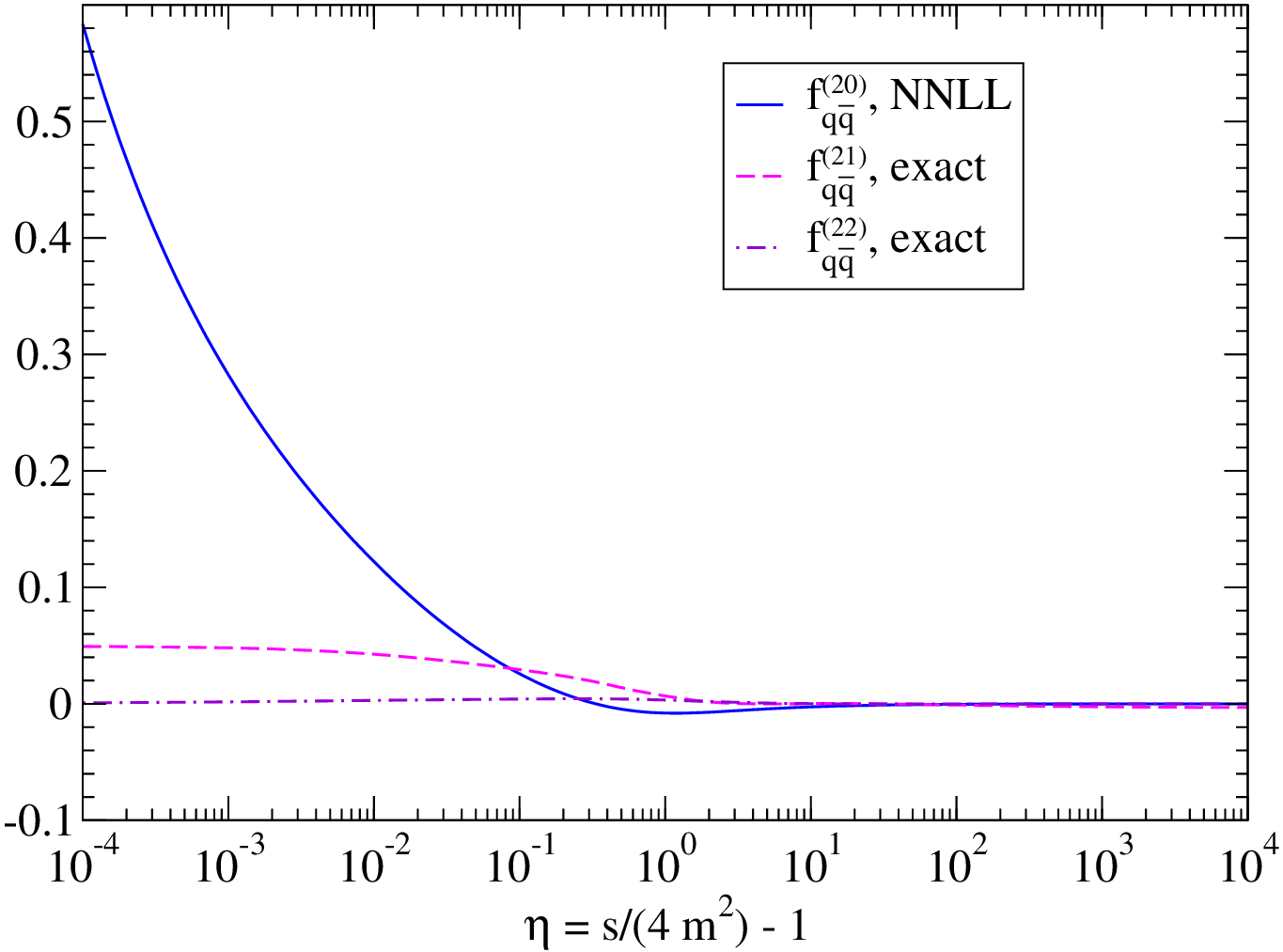}}
    }
\hspace*{3mm}
\subfigure[\label{subfig:fq1q22}]
    {
        \scalebox{0.32}{\includegraphics{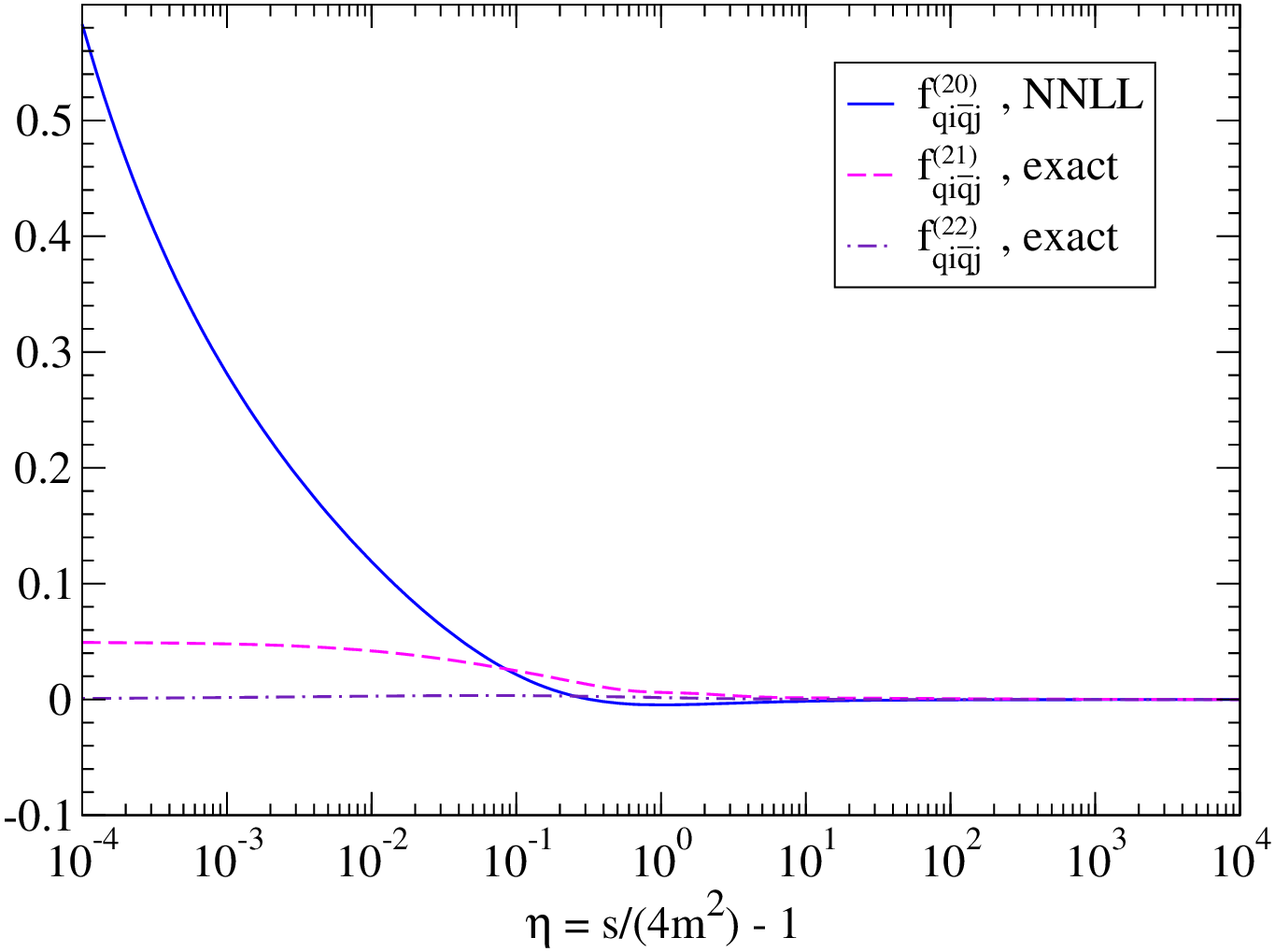}}
    }
\vspace*{6mm}

\subfigure[\label{subfig:fqq3}]
    {
        \scalebox{0.32}{\includegraphics{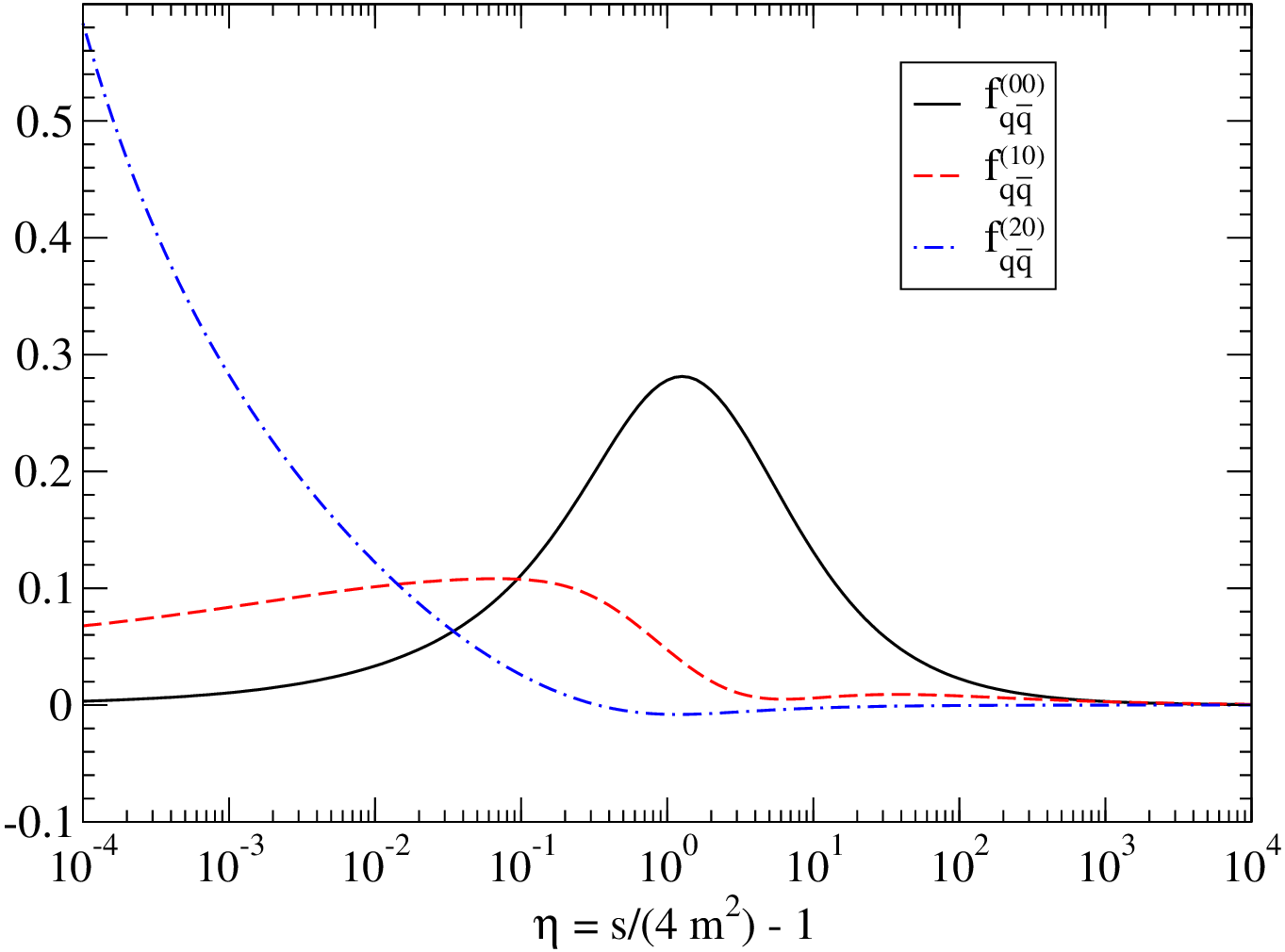}}
    }
\hspace*{3mm}
\subfigure[\label{subfig:fq1q23}]
    {
        \scalebox{0.32}{\includegraphics{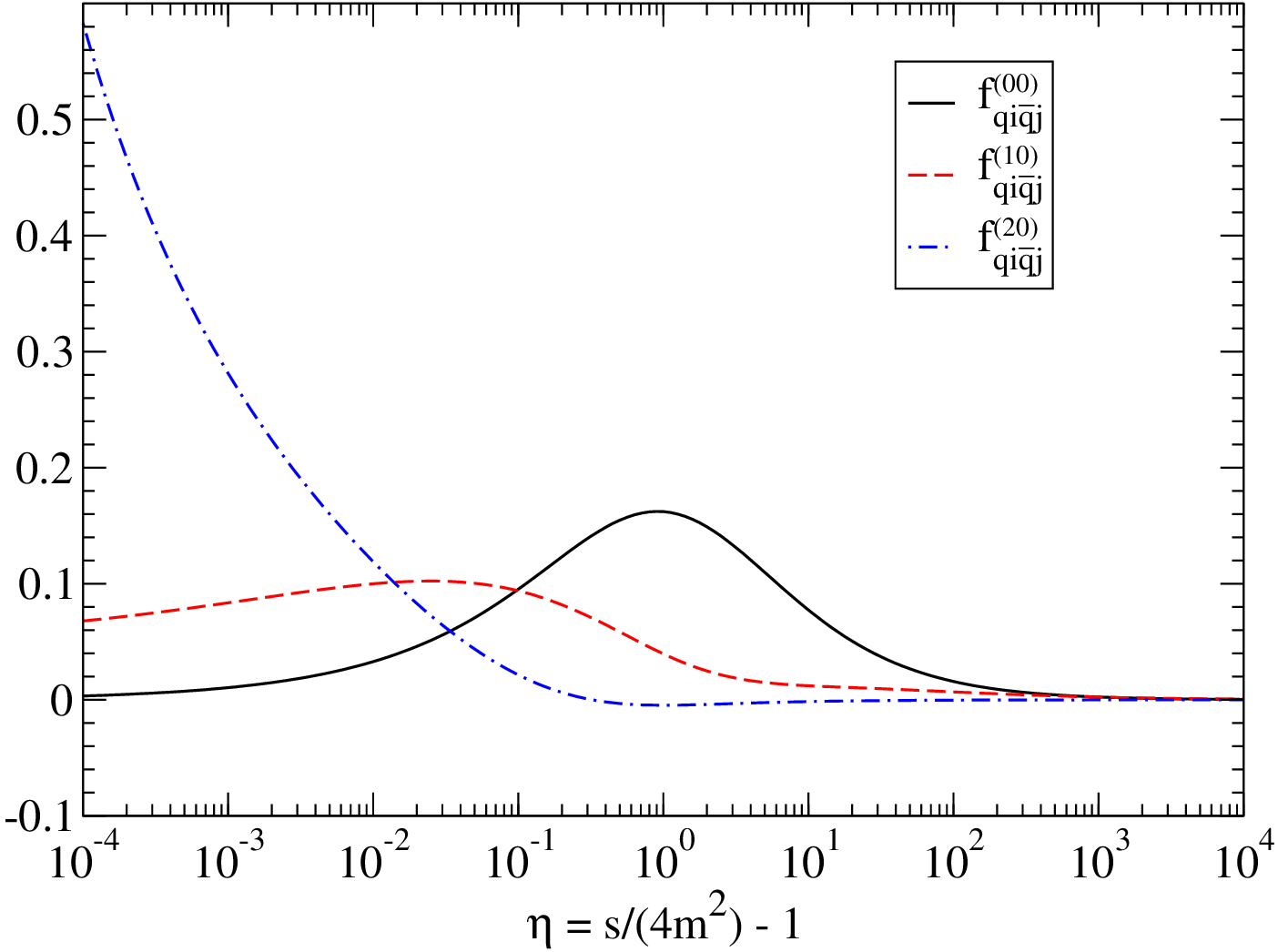}}
    }
\caption{\small
\label{fig:fqqfunction}
  Scaling functions $f_{q\barq}^{(ij)}$ (left column) and $f_{q_k \barq_l}^{(ij)}$ 
  (right column) with $i = 0\ldots 2,\enspace j\le i$, for $\mq = 400\GeV$ and 
  $\mg = 500\GeV$.}
\end{figure}
% \newpage

\begin{figure}
\centering
\subfigure[\label{subfig:fgg1}]
    {
        \scalebox{0.33}{\includegraphics{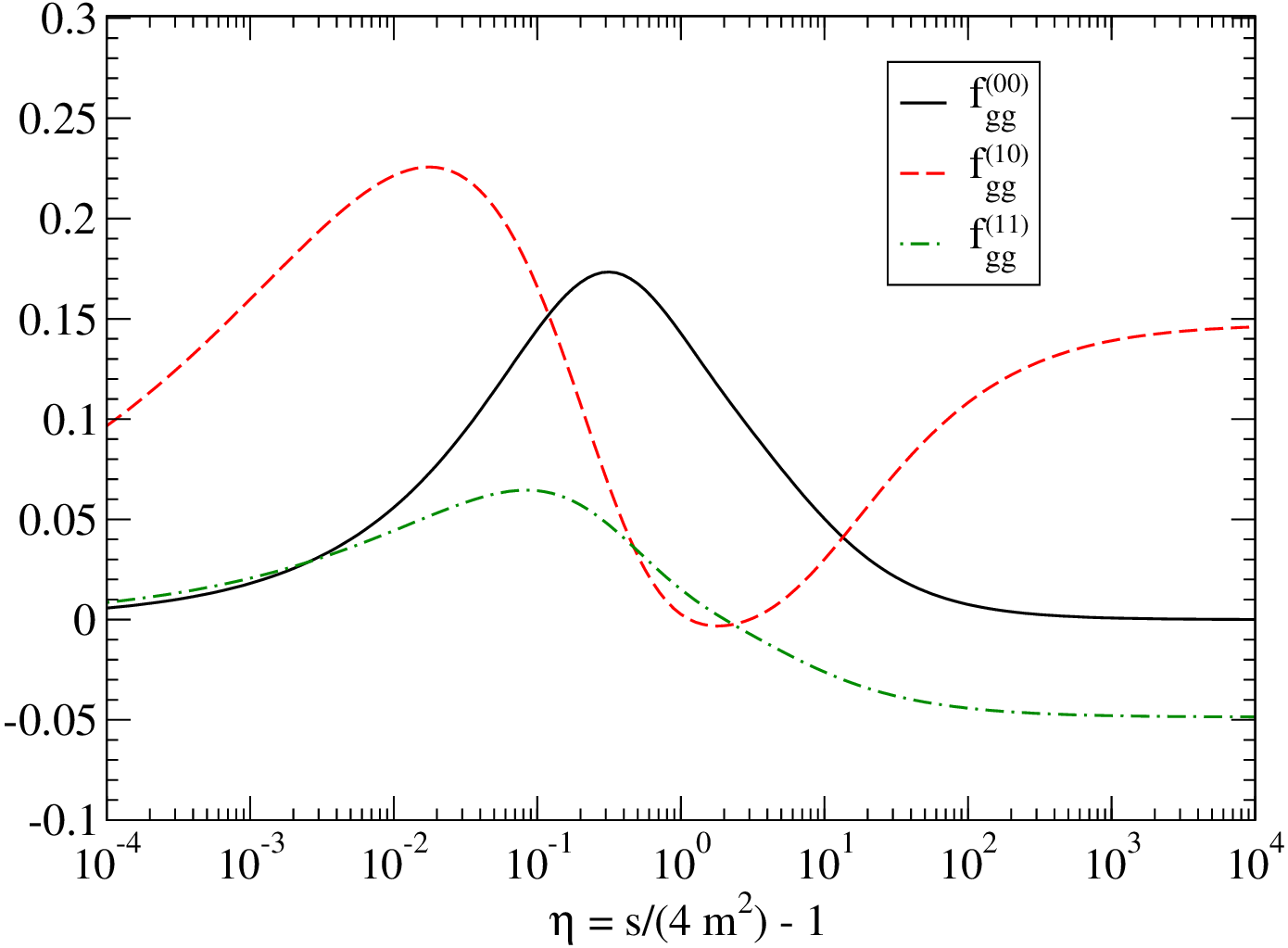}}
    }
\vspace*{6mm}

\subfigure[\label{subfig:fgg2}]
    {
        \scalebox{0.32}{\includegraphics{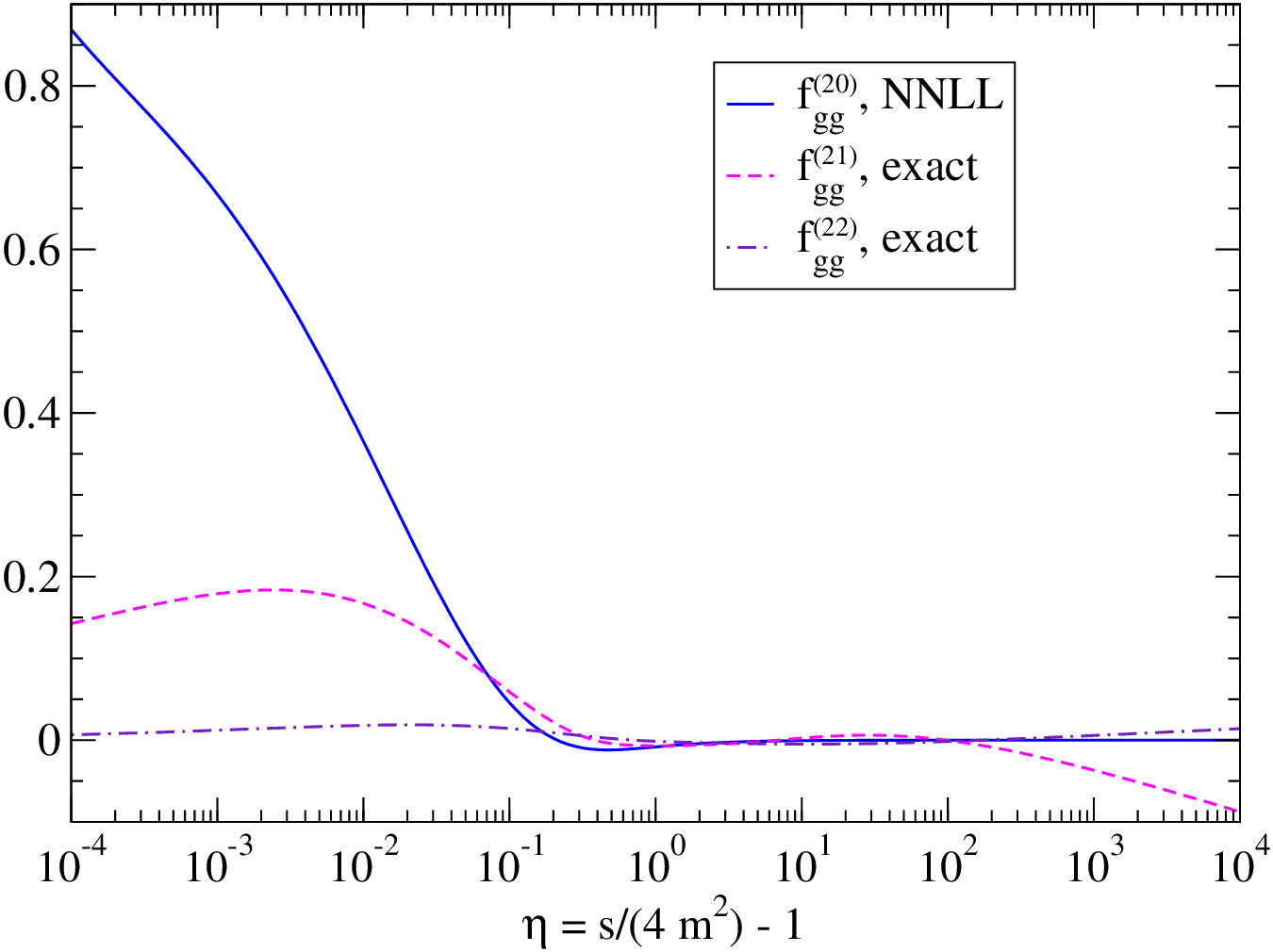}}
    }
\vspace*{6mm}

\subfigure[\label{subfig:fgg3}]
    {
        \scalebox{0.33}{\includegraphics{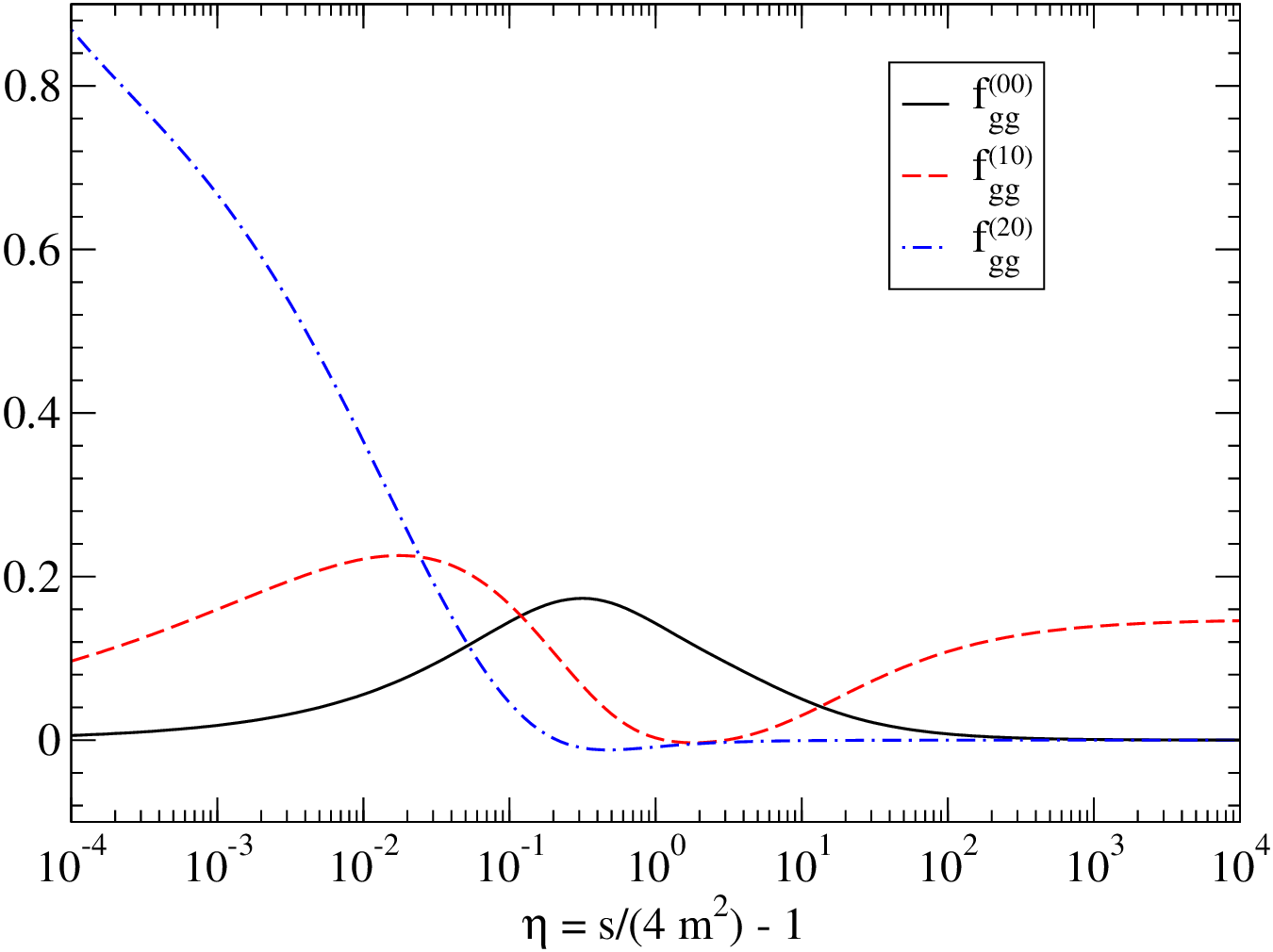}}
    }
\caption{\small
\label{fig:fggfunction}
  Scaling functions $f_{gg}^{(ij)}$ with $i = 0,1,2\enspace j\le i$.}
\end{figure}
Let us start to illustrate the phenomenological consequences. 
In our numerical illustration we assume throughout this paper 
for the gluino mass the relation $\mg = 1.25 \*\mq$.
Here our choice of parameters $\mq = 400\GeV$ and $\mg = 500\GeV$ as a
reference point has been influenced by the current limits from direct searches. 
Thus far, lower limits on $\mq$ and $\mg$ are provided by Tevatron 
(CDF and D0 collaboration~\cite{Peters:2008fn,:2007ww,Adams:2008nu}) 
and an absolute lower limit of $379\GeV$ and $308\GeV$, respectively, 
for the masses of squarks and the gluino in the mSUGRA framework ($A_0 = 0, \mu < 0, \tan\beta = 3$) 
has been quoted.
Moreover, the non-stop-squarks are expected to be in a narrow mass range, because
the Standard Model quarks are (nearly) massless and, 
in this case, nearly mass-degenerate squarks are a property of mSUGRA.
In the mSUGRA framework, the $\tilde{t}_1$ is lighter and the $\tilde{t}_2$ heavier 
than the other squarks and a typical particle spectrum can be found in
Ref.~\cite{AguilarSaavedra:2005pw}. 
As announced already above, we have excluded stop quarks from our considerations here. 
Currently, the NLO QCD corrections to stop production are known~\cite{Beenakker:1997ut} 
(see e.g. Ref.~\cite{Abazov:2008kz} for a search for the lightest stop $\tilde{t}_1$).

We display in Figs.~\ref{fig:fqqfunction} and \ref{fig:fggfunction} our results for the 
scaling functions as defined in Eq.~(\ref{eq:ggkl}) up to second order in $\as$ and separated
according to the parton initial state $gg$ and $q\bar{q}$.
The results for $f_{ij}^{(20)}$, $f_{ij}^{(21)}$ and $f_{ij}^{(22)}$ are new.
To illustrate the effect of threshold logarithms, we compare the scale independent functions 
$f_{ij}^{(00)}$, $f_{ij}^{(10)}$ and $f_{ij}^{(20)}$ directly 
in Figs.~\ref{subfig:fqq3}, \ref{subfig:fq1q23} and \ref{subfig:fgg3}.
One clearly sees the effect of the large logarithms in $\beta$ at low $\eta$
giving rise to large perturbative corrections.
At larger $\eta$, the results for $f_{ij}^{(20)}$ from
Eqs.~(\ref{eq:fgg20-num})--(\ref{eq:fq1q220-num}) 
vanish quickly and our approximation ceases to be valid for $\eta \gsim {\cal O}(1 \dots 10)$.

\bigskip

\begin{figure}[ht!]
\centering
\vspace*{10mm}
\subfigure%[\label{subfig:lhcnlo}]
    {
        \scalebox{0.32}{\includegraphics{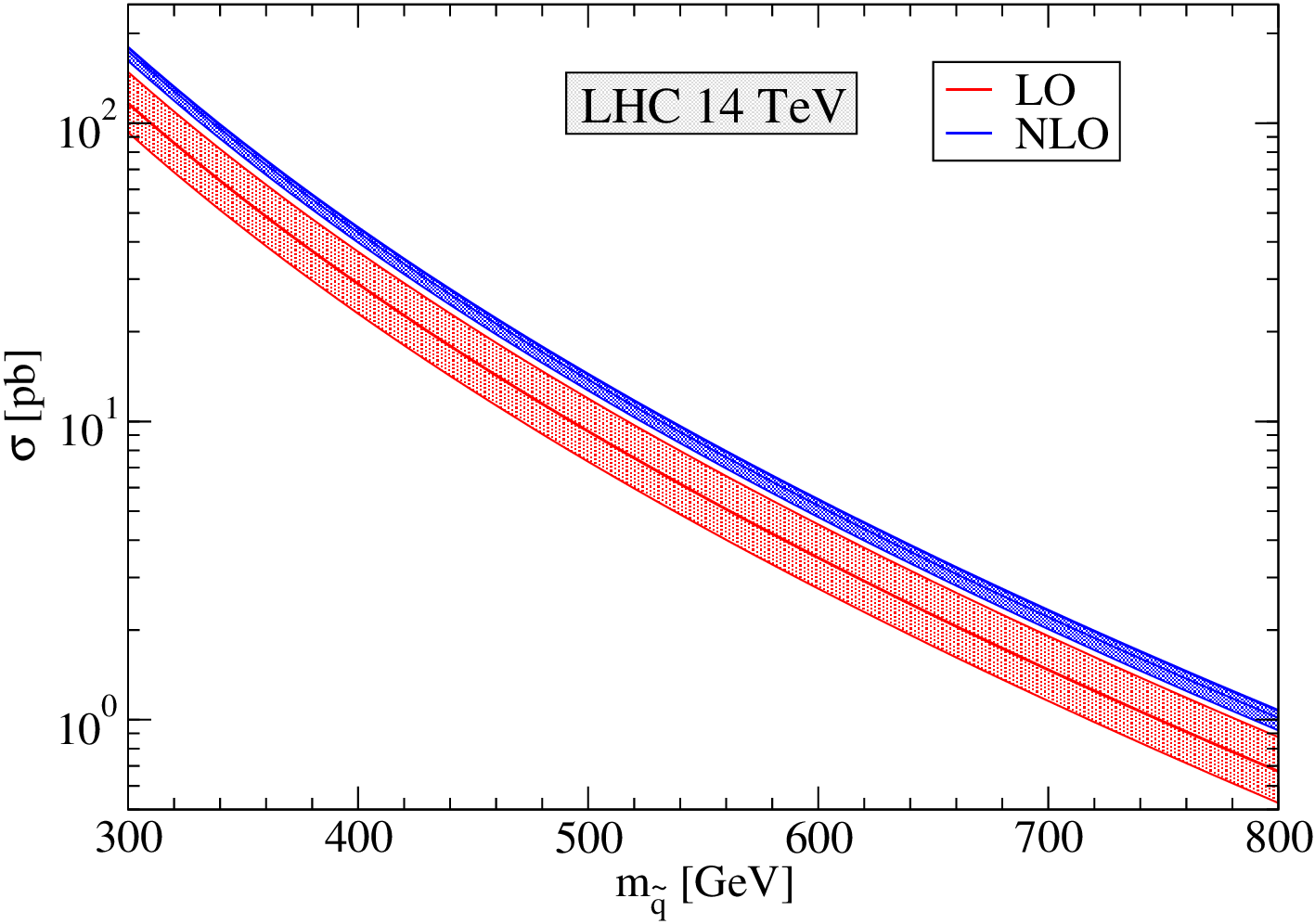}}
    }
% \hspace*{2mm}
\subfigure%[\label{subfig:lhcnnlo}]
    {
        \scalebox{0.32}{\includegraphics{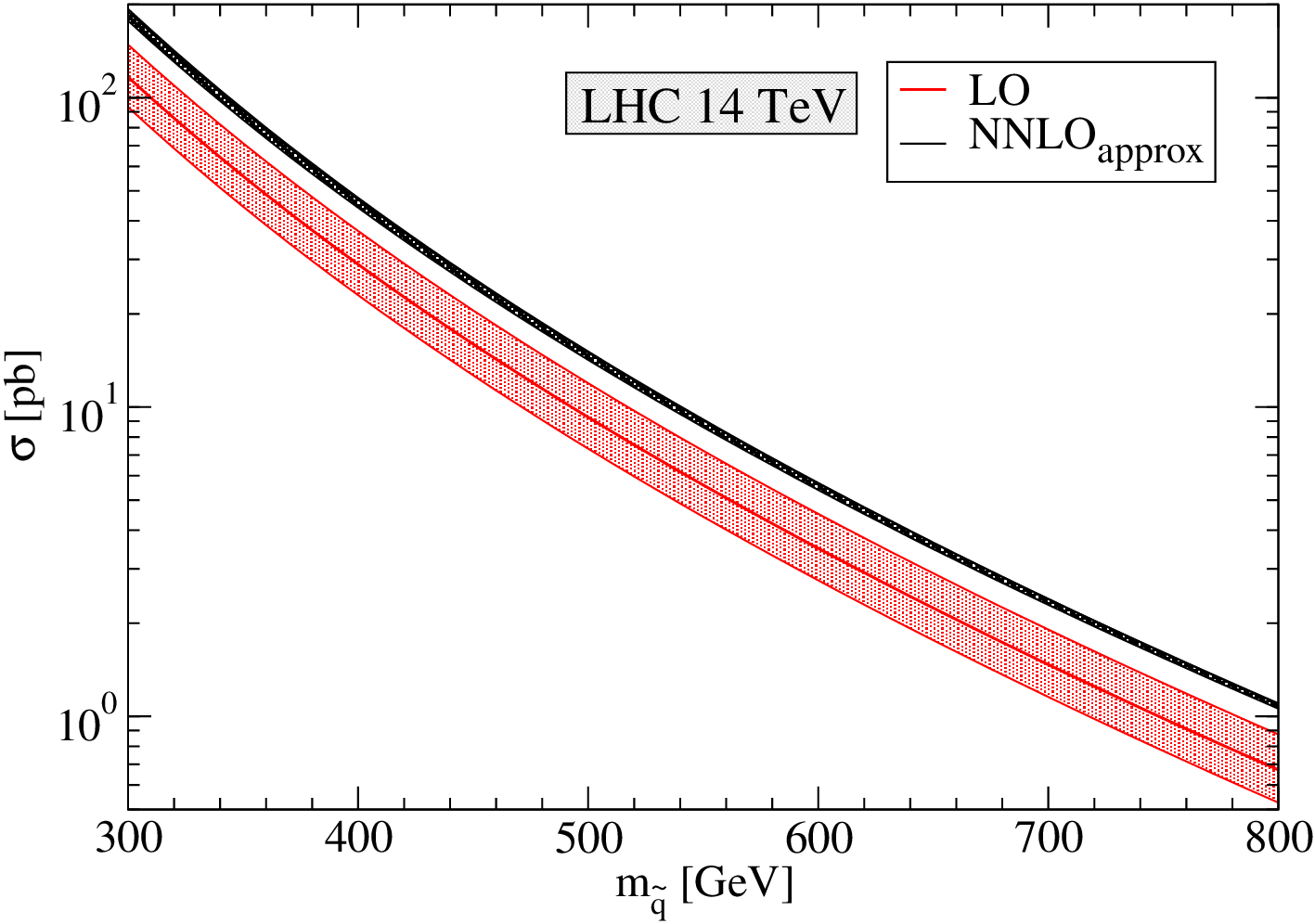}}
    }
\caption{\small
  \label{fig:lhc14}
  The total hadronic cross section for 
  $pp\to \sq \sqb$ at LHC with $\sqrt{s}=14~\TeV$  
  to LO, NLO, and NNLO in QCD as function of the squark mass $\mq$.
  For the gluino mass we assume the relation $\mg = 1.25 \mq$.}
\end{figure}
\begin{figure}[ht!]
 \centering
 \vspace*{5mm}
 \subfigure%[\label{subfig:tevanlo}]
    {
        \scalebox{0.32}{\includegraphics{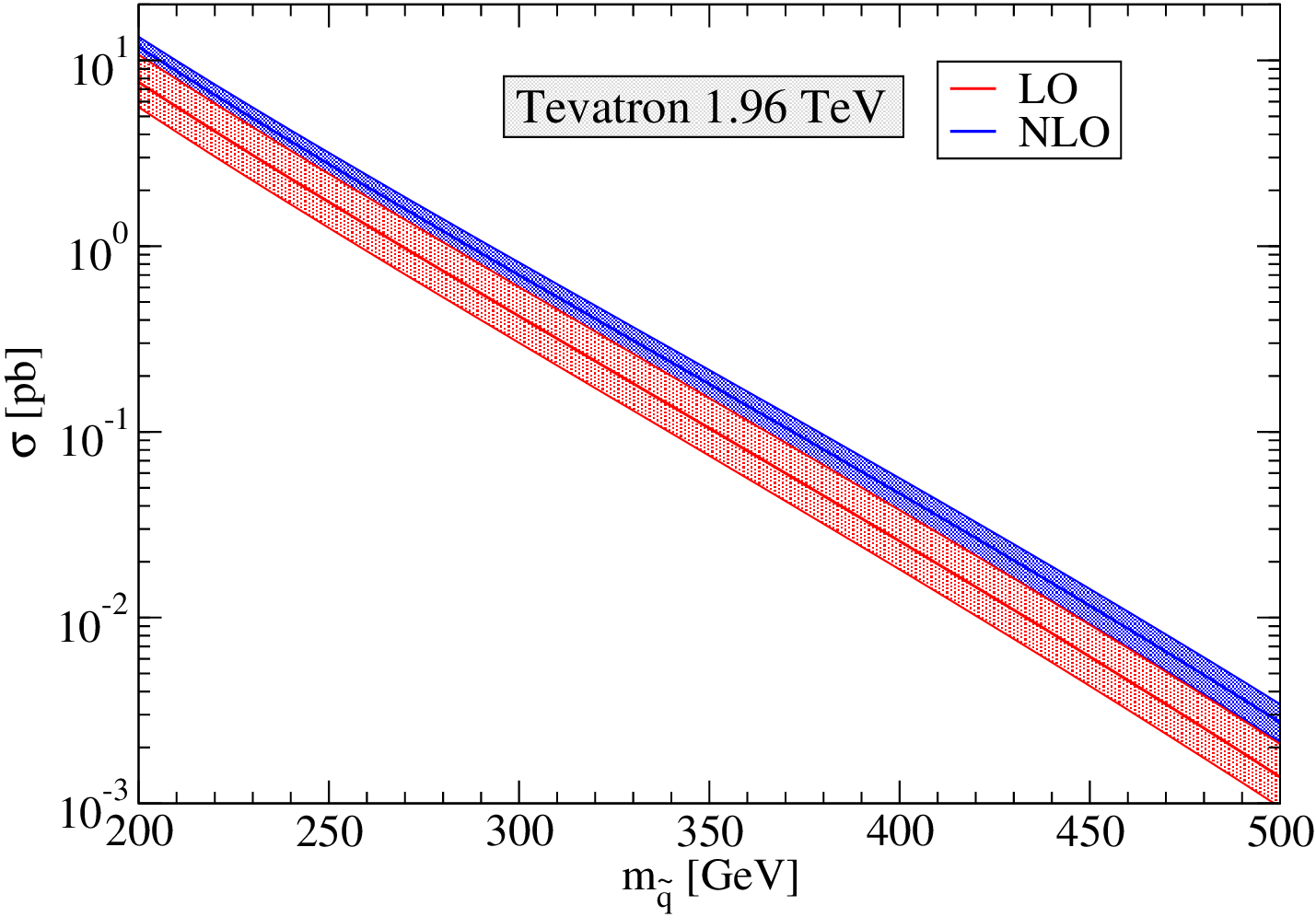}}
    }
% \hspace*{2mm}
\subfigure%[\label{subfig:tevannlo}]
    {
        \scalebox{0.32}{\includegraphics{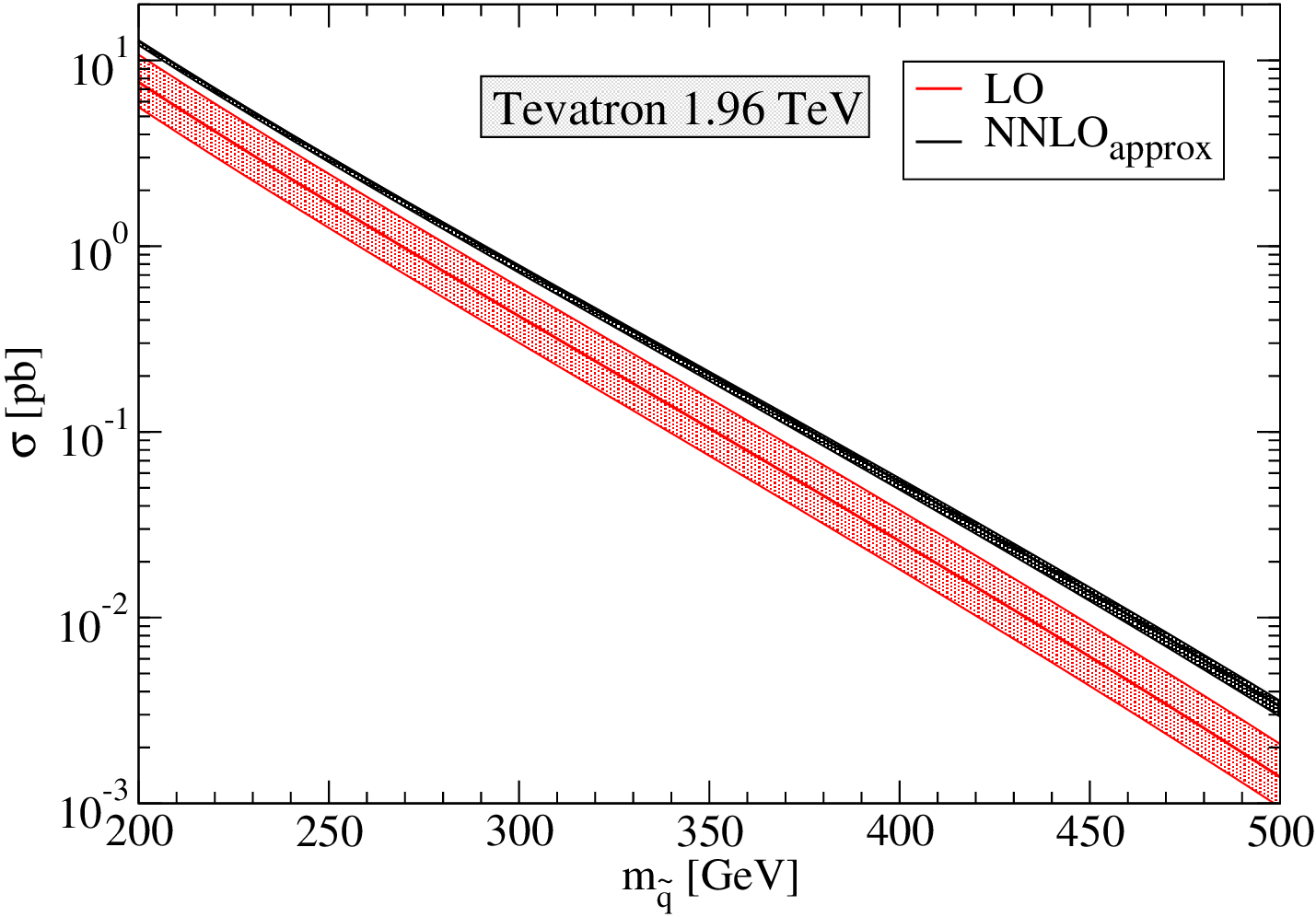}}
    }
\caption{\small
  \label{fig:teva}
  Same as Fig.~\ref{fig:lhc14} for Tevatron with $\sqrt{s}=1.96~\TeV$.
  }
\end{figure}
Let us next investigate the consequences for the hadronic cross section~(\ref{eq:totalcrs}) 
for squark pair-production at LHC and Tevatron. 
The necessary convolution of the parton scaling functions with the parton
luminosity~(\ref{eq:partonlumi}) emphasizes the threshold region of phase space.
The parton luminosity $L_{ij}$ is steeply falling with increasing energies.
Consequently, the total hadronic cross section is effectively saturated from partonic 
processes close to threshold 
with the kinematics being very similar to the case of top-quark hadro-production~\cite{Moch:2008qy}.
As an upshot, one can conclude, that our approximate NNLO result captures the numerically dominant 
part of the complete (yet unknown) NNLO corrections.
Eqs.~(\ref{eq:fgg20-num})--(\ref{eq:fq1q220-num}) should thus represent a very
reliable estimate.

In Fig.~\ref{fig:lhc14}, we present the total cross section at the LHC as a
function of the squark mass $\mq$.
We use the PDF set CTEQ6.6~\cite{Nadolsky:2008zw}, if not stated otherwise.
In the left figure, we show the LO and NLO QCD predictions for the cross
sections along with their error bands due to the scale uncertainty for scale choices 
$x \equiv \mu/\mq = 2$ and $x = 1/2$. 
In Fig.~\ref{fig:lhc14} on the right, we show the same comparison for the LO and NNLO prediction.
While the error band of the LO cross section is quite large 
(due to the scale uncertainty of the strong coupling constant $\as$ and the PDFs)
it shrinks when considering the NLO QCD prediction.
For the cross section at NNLO accuracy, we apply our newly derived scaling
functions $f_{ij}^{(20)}$, $f_{ij}^{(21)}$ and $f_{ij}^{(22)}$ of the previous
Section and observe how the error band contracts to a thin line. 
This proves the significant reduction in scale uncertainty at this order of perturbation theory.
Numbers for the hadronic cross section for different squark masses and scales
are listed in Tab.~\ref{tab:xsecvalues}.
The NLO cross section increases by about $50\%$ compared to the LO cross section,
and likewise, the approximate NNLO cross section by about $9\%$ compared to the NLO result.
The latter value, i.e. $9\%$, is to be contrasted with the rather small $K$-factor 
for the NLL resummed cross section reported in Ref.~\cite{Kulesza:2008jb} of
about $1 - 2\%$ in the range $\mq = 200$--$1000\GeV$. 
As a direct phenomenological consequence we can translate our result into
shifts of the exclusion limits for the squark masses for one example.
For instance, a cross section of $30\pb$ corresponds at LO, NLO, and NNLO to a squark 
mass of about $397\GeV$, $430\GeV$, and $437\GeV$, respectively.

In Fig.~\ref{fig:teva}, we present the same plot for the total cross section
of $p{\bar p}\to \sq\sqb$ at the Tevatron.
Here, the cross section is more than two orders of magnitude smaller than at LHC 
due to the lower center-of-mass energy of Tevatron.
Again, we observe improved stability of the perturbative predictions with
respect to the scale variation.
In Tab.~\ref{tab:xsectevavalues}, we give explicit values for the cross
section and the NNLO result implies new exclusion limits. For example 
for a cross section of $60\fb$ one
comes up with a squark mass of about $370\GeV$, $391\GeV$ and $396\GeV$
at LO, NLO, and NNLO, respectively.

\begin{figure}[ht!]
 \centering
 \vspace*{5mm}
 \subfigure[\label{fig:ggqqlhc}]
    {
        \scalebox{0.32}{\includegraphics{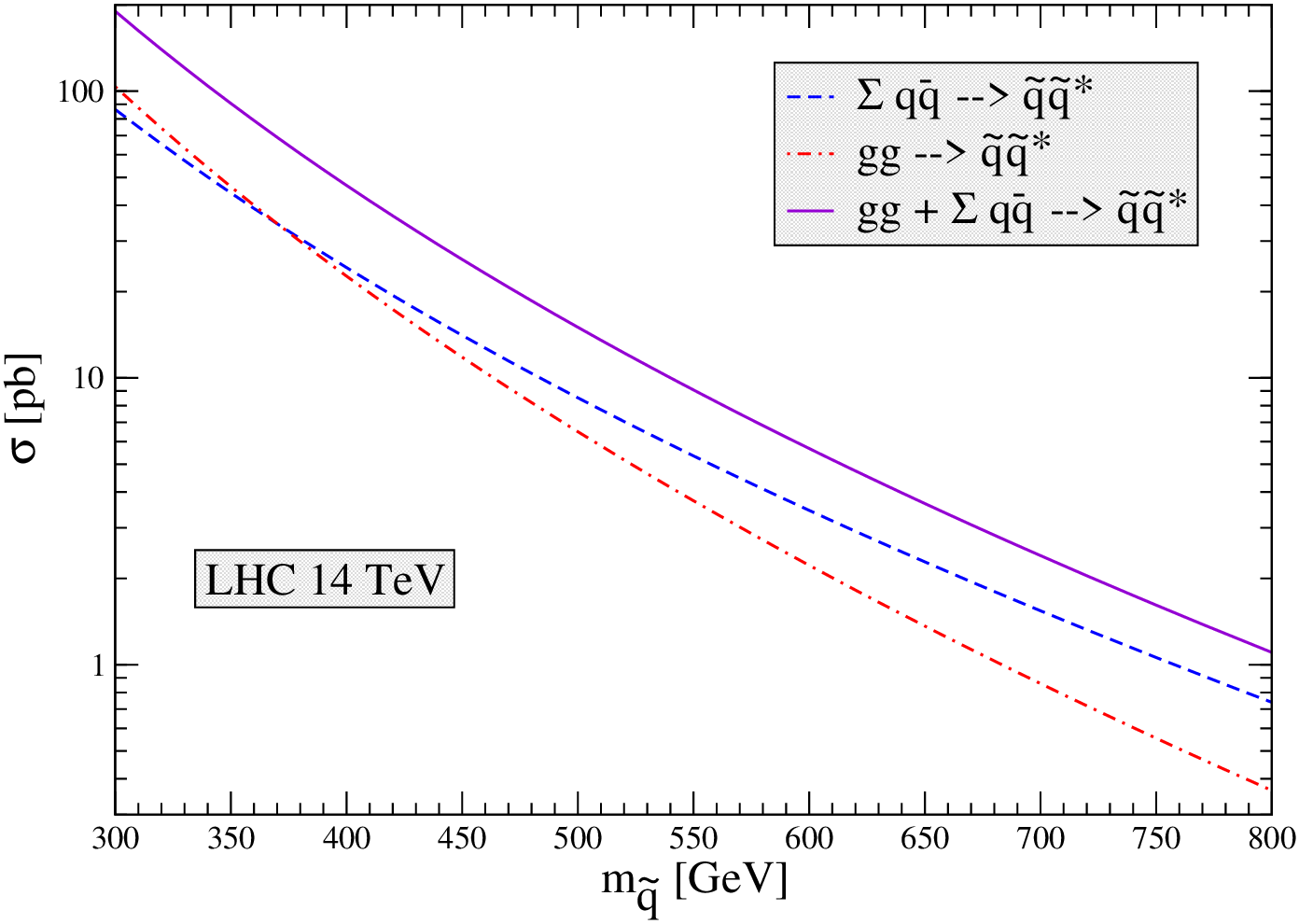}}
    }
% \hspace*{2mm}
\subfigure[\label{fig:ggqqteva}]
    {
        \scalebox{0.32}{\includegraphics{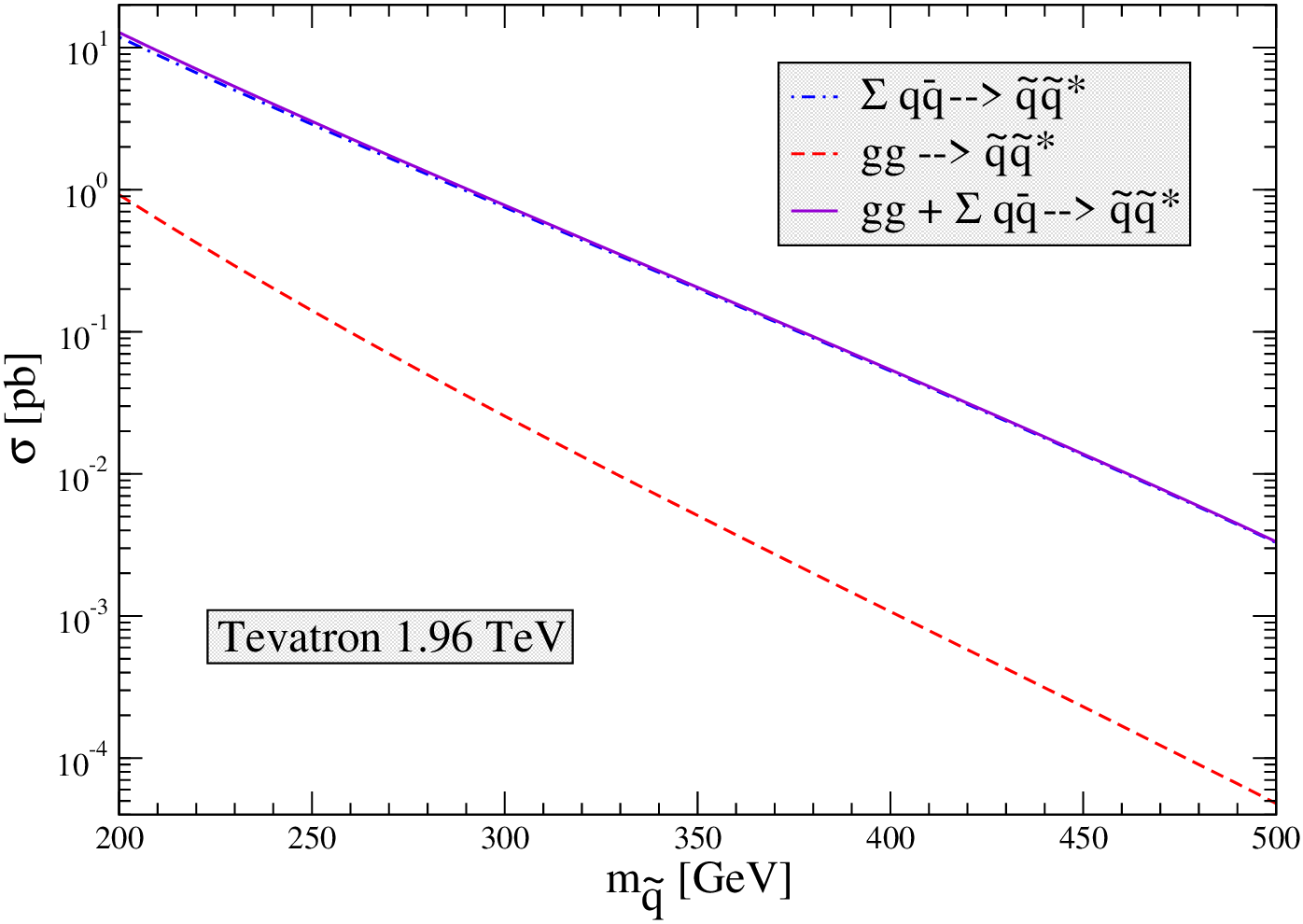}}
    }
\caption{\small
 \label{fig:tot}
The contributions of the channels $gg \to \sq \sqb$ and $\sum q \barq \to \sq \sqb$ 
to the total hadronic cross section at NNLO for the LHC (left figure) and the Tevatron (right figure).
The gluino mass is $\mg = 1.25 \mq$ and the scale $\mu = \mq$.
}
\end{figure}

In Fig.~\ref{fig:tot}, we show the contributions of the different parton 
channels~(\ref{eq:Borngg}) and (\ref{eq:Bornqq}) to the total cross section.
At the LHC, the contribution of the gluon channel is dominant only up to a squark
mass of about $370\GeV$. 
The ratio of the two channels, $\sigma(\sum q\barq \to \sq\sqb)/\sigma(gg \to \sq\sqb)$,
changes from $45\% / 55\%$ for $\mq=300\GeV$ to $67\% / 33\%$ for $\mq=800\GeV$.
This is not what one naively expects from a $pp$-collider.
The reason for this behavior is the summation over all possible initial quark flavors
and final squark flavors and chiralities. The gluon channel has only $10$
contributing subprocesses, the quark channel however $140$ subprocesses.
At the Tevatron, the gluon channel is nearly negligible as one can see from 
Fig.~\ref{fig:ggqqteva} and the large number 
of contributing subprocesses in the quark channel is additionally enhanced 
by the large $q{\bar q}$-luminosity.
The ratio of the two channels increases from $93\% / 7\%$ for $\mq=200\GeV$ 
to $98.5\% / 1.5\%$ for $\mq=500\GeV$.

\begin{figure}[ht!]
 \centering
 \vspace*{5mm}
 \subfigure[\label{fig:scale}]
    {
        \scalebox{0.32}{\includegraphics{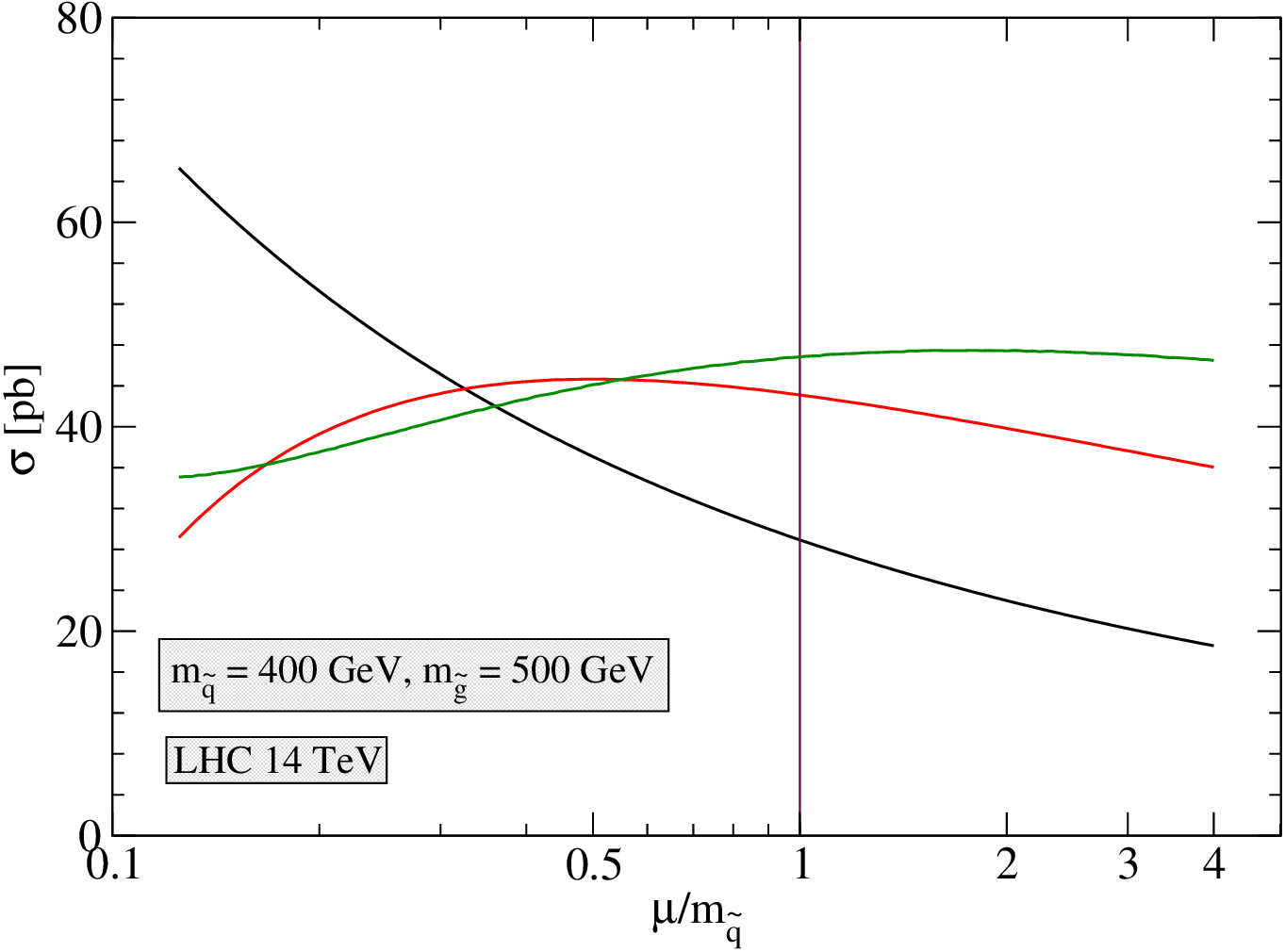}}
    }
% \hspace*{2mm}
\subfigure[\label{fig:slhc}]
    {
        \scalebox{0.32}{\includegraphics{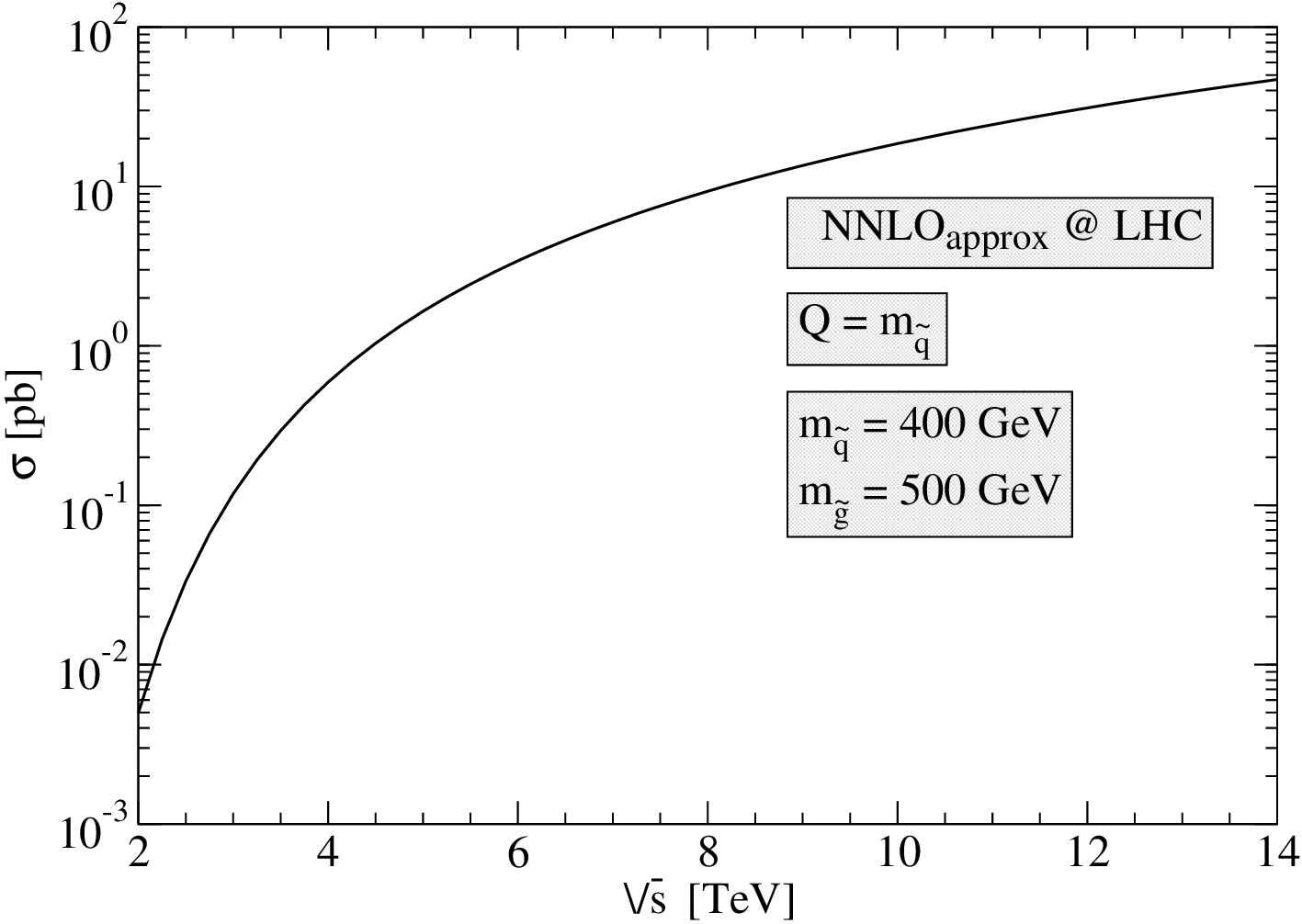}}
    }
\caption{\small
 \label{fig:many}
 The total hadronic cross section for 
 $pp\to \sq \sqb$ at LHC as a function of 
 the scale $\mu$ (left figure) and as a function of 
 the center-of-mass energy $\sqrt{s}$ (right figure) 
 for the reference point.
 On the left hand side we compare the LO, NLO, and NNLO predictions for the
 $\mu$ dependence with $\sqrt{s}=14~\TeV$.
 On the right we display the $\sqrt{s}$ dependence of our NNLO prediction at
 scale $\mu = \mq$.
}
\end{figure}
\begin{figure}[ht!]
\centering
\vspace*{5mm}
\subfigure%[\label{subfig:ctq66err}]
    {
        \scalebox{0.32}{\includegraphics{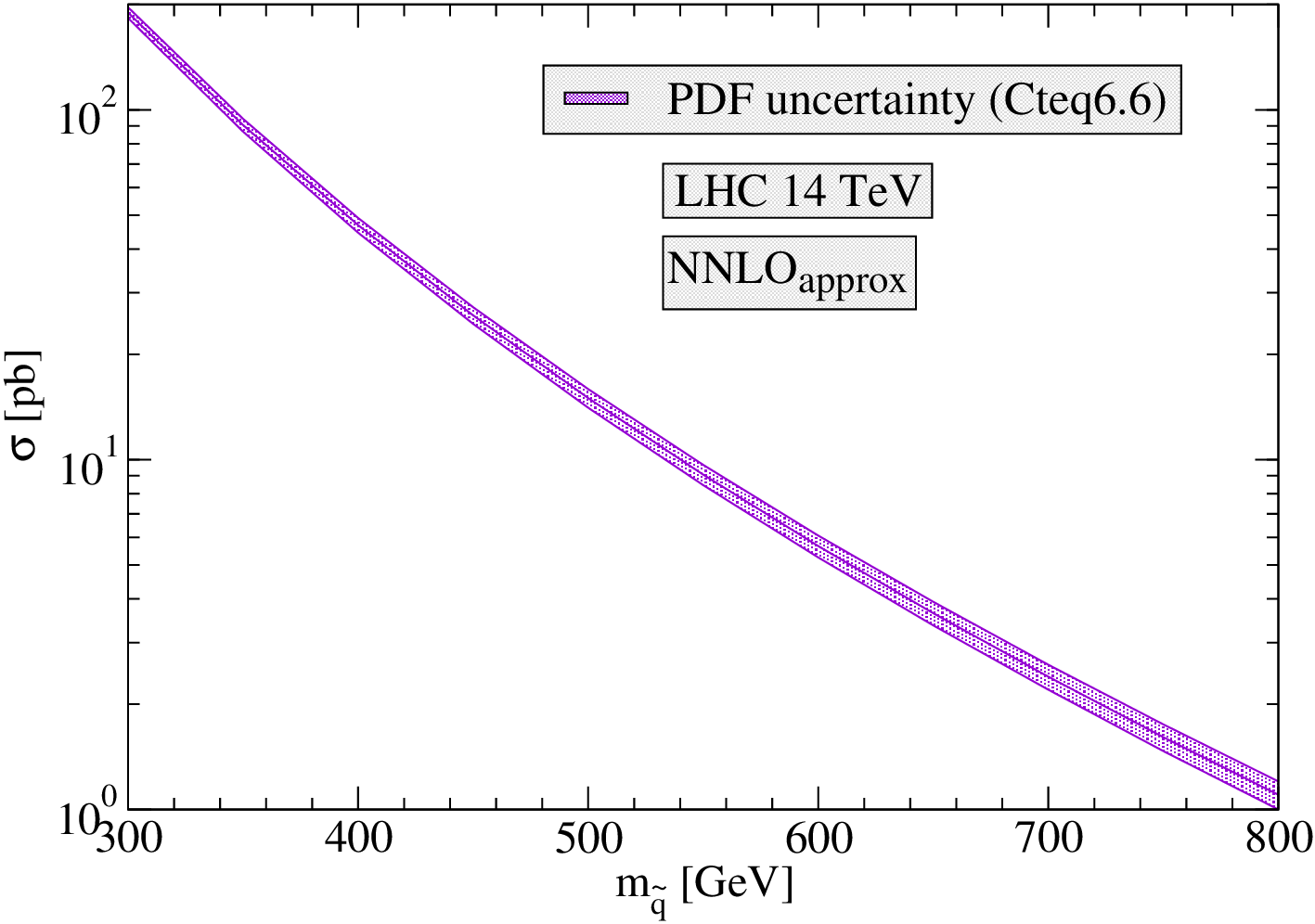}}
    }
% \hspace*{2mm}
\subfigure%[\label{subfig:mrsterr}]
    {
        \scalebox{0.32}{\includegraphics{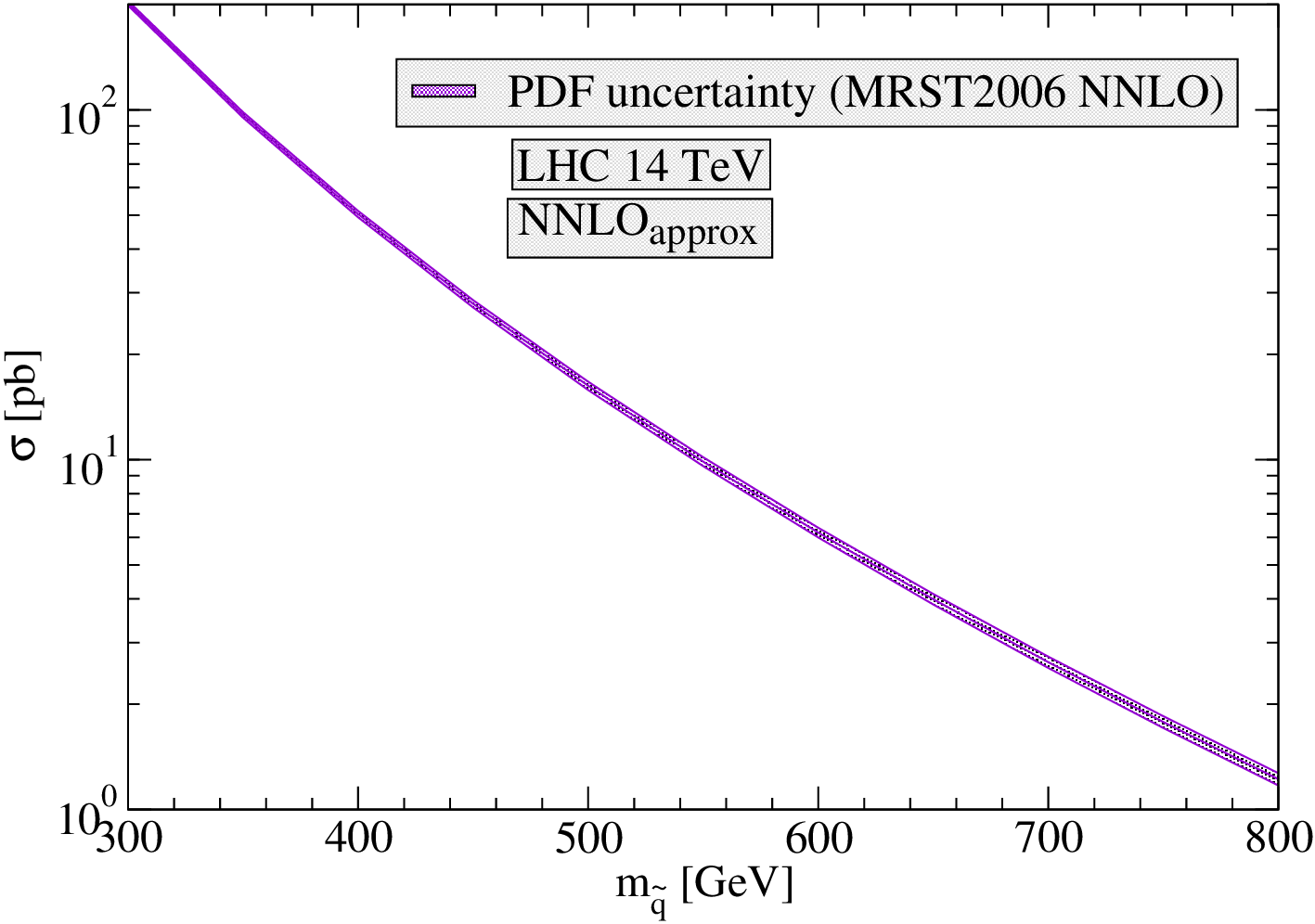}}
    }
\caption{\small
  \label{fig:errorpdf}
  The PDF uncertainty of the NNLO cross section for the two PDF sets
  CTEQ6.6~\cite{Nadolsky:2008zw} (left figure) and MRST2006nnlo~\cite{Martin:2007bv} (right figure).}
\end{figure}
In order to illustrate the improved theoretical prediction due to the reduced
scale dependence, we show in Fig.~\ref{fig:scale} 
the LO, NLO, and NNLO predictions for the cross section at LHC with $\sqrt{s}=14~\TeV$.
One observes a decreasing dependence on $\mu$ with increasing order of perturbation theory. 
Within the commonly chosen scale interval $x=0.5 \ldots 2$, the
LO, NLO, and NLO cross section varies between $37\pb$ and $23\pb$, $45\pb$ and $40\pb$,
and $44\pb$ and $47\pb$, respectively (see Tab.~\ref{tab:xsecvalues}).
This amounts to a residual theoretical uncertainty of only ${\pm} 3\%$ for our
approximate NNLO prediction 
and improves the value for the resummed cross section to NLL accuracy 
of about ${\pm} 10\%$ quoted in Ref.~\cite{Kulesza:2008jb}.

The presently discussed schedule for the initial phase of LHC
includes operation at a center-of-mass energy lower than $14\TeV$, 
a value of $10\TeV$ has often been quoted.
For this reason, it is interesting to study the energy dependence of the cross section.
In Fig.~\ref{fig:slhc} we display our new approximate NNLO result 
as a function of the center-of-mass energy.
We observe, for instance, at $10\TeV$ a cross section of about $19\pb$ for our
reference point ($\mq = 400\GeV$ and $\mg = 500\GeV$)
compared to $47\pb$ for the design energy of $14\TeV$.

The final point of interest is the sensitivity to the parton luminosity and
the associated uncertainty. 
In Fig.~\ref{fig:errorpdf}, we show the NNLO cross section for squark pair-production
together with the error band due to the PDF uncertainties.
We used two sets of PDFs: CTEQ6.6 (left figure,~\cite{Nadolsky:2008zw}) 
and MRST2006nnlo (right figure,~\cite{Martin:2007bv}).
The PDF uncertainty has been calculated accordingly to 
Ref.~\cite{Nadolsky:2008zw}.
At higher squark masses, the error on the cross section is significantly increased 
because the relevant phase space probes the gluon luminosity in the high-$x$ 
region where the gluon PDF has a large uncertainty.
The relative error increases from $\pm 3.5\%$ at $\mq = 300\GeV$
to $\pm 9\%$ at $\mq = 800\GeV$.
Note that the quoted error for the MRST2006nnlo PDF set is significantly smaller 
than for the CTEQ6.6 set. 
Its error increases from $\pm 1.5\%$ to $\pm 4\%$ for squark masses ranging form
$300\GeV$ to $800\GeV$.
\begin{table}
\centering
 \begin{tabular}{r|rrr|rrr|rrr}
\toprule
$\mq$ &\multicolumn{3}{c|}{$\sigma(\text{LO}) [\pb]$}&
       \multicolumn{3}{c|}{$\sigma(\text{NLO}) [\pb]$}&
       \multicolumn{3}{c}{$\sigma(\text{NNLO}) [\pb]$}\\
$[\GeV]$& $x = \half$& $x = 1$& $x = 2$& $x = \half$& $x = 1$& $x = 2$& $x = \half$& $x = 1$& $x = 2$\\
\hline  
$  300 $ & $  148.4 \phantom{0} $ &  $  116.6 \phantom{0} $ &  $   93.2 \phantom{0} $ & $  179.5 \phantom{0} $ &  $  174.2 \phantom{0} $ &  $  161.9 \phantom{0} $ & $  177.4 \phantom{0} $ &  $  189.8 \phantom{0} $ &  $  193.4 \phantom{0} $  \\[1mm]
$  400 $ & $   37.1 \phantom{0} $ &  $   28.9 \phantom{0} $ &  $   23.0 \phantom{0} $ & $   44.6 \phantom{0} $ &  $   43.1 \phantom{0} $ &  $   39.8 \phantom{0} $ & $   44.1 \phantom{0} $ &  $   46.8 \phantom{0} $ &  $   47.4 \phantom{0} $  \\[1mm]
$  500 $ & $   11.9 \phantom{0} $ &  $    9.3 \phantom{0} $ &  $    7.3 \phantom{0} $ & $   14.4 \phantom{0} $ &  $   13.8 \phantom{0} $ &  $   12.7 \phantom{0} $ & $   14.2 \phantom{0} $ &  $   15.0 \phantom{0} $ &  $   15.1 \phantom{0} $  \\[1mm]
$  600 $ & $   4.51 $ &  $   3.49 $ &  $   2.75 $ & $   5.47 $ &  $   5.22 $ &  $   4.78 $ & $   5.39 $ &  $   5.66 $ &  $   5.68 $  \\[1mm]
$  700 $ & $   1.91 $ &  $   1.47 $ &  $   1.16 $ & $   2.33 $ &  $   2.21 $ &  $   2.01 $ & $   2.29 $ &  $   2.40 $ &  $   2.40 $  \\[1mm]
$  800 $ & $   0.87 $ &  $   0.67 $ &  $   0.53 $ & $   1.07 $ &  $   1.01 $ &  $   0.92 $ & $   1.06 $ &  $   1.10 $ &  $   1.10 $  \\[1mm]
$  900 $ & $   0.43 $ &  $   0.33 $ &  $   0.25 $ & $   0.53 $ &  $   0.50 $ &  $   0.45 $ & $   0.52 $ &  $   0.54 $ &  $   0.53 $  \\[1mm]
$ 1000 $ & $   0.22 $ &  $   0.17 $ &  $   0.13 $ & $   0.27 $ &  $   0.25 $ &  $   0.23 $ & $   0.27 $ &  $   0.28 $ &  $   0.27 $  \\[1mm]
\bottomrule
 \end{tabular}
    \caption{\small
      \label{tab:xsecvalues}
      Numerical values for the squark pair-production cross section 
      at LHC with $\sqrt{s}=14\TeV$ and the CTEQ6.6 PDF set~\cite{Nadolsky:2008zw}.
      The QCD predictions are given at LO, NLO, and NNLO accuracy and  
      for different squark masses and scales $x = \mu/\mq$.}
\end{table}
\begin{table}
\centering
 \begin{tabular}{>{\hspace{-2mm}}r|>{\hspace{-2pt}}r>{\hspace{-3pt}}r>{\hspace{-2pt}}r|
                                   >{\hspace{-2pt}}r>{\hspace{-3pt}}r>{\hspace{-2pt}}r|
                                   >{\hspace{-2pt}}r>{\hspace{-3pt}}r>{\hspace{-2pt}}r}
\toprule
$\mq$ &\multicolumn{3}{c|}{$\sigma(\text{LO}) [\fb]$}&
       \multicolumn{3}{c|}{$\sigma(\text{NLO}) [\fb]$}&
       \multicolumn{3}{c}{$\sigma(\text{NNLO}) [\fb]$}\\
$[\GeV]$& $x = \half$& $x = 1$& $x = 2$& $x = \half$& $x = 1$& $x = 2$& $x = \half$& $x = 1$& $x = 2$\\
\hline
$  200 $ & $ 10735.0   $ &  $ 7641.1   $ &  $ 5606.6   $ & $ 13394.7   $ &  $ 11836.7   $ &  $ 10091.7   $ & $ 12707.5   $ &  $ 12749.7   $ &  $ 12100.4   $  \\[1mm]
$  250 $ & $ 2450.5   $ &  $ 1726.6   $ &  $ 1254.3   $ & $ 3182.5   $ &  $ 2766.6   $ &  $ 2328.8   $ & $ 3043.4   $ &  $ 3019.7   $ &  $ 2835.0   $  \\[1mm]
$  300 $ & $  602.4   $ &  $  419.9   $ &  $  301.9   $ & $  816.9   $ &  $  698.6   $ &  $  580.4   $ & $  789.9   $ &  $  775.2   $ &  $  720.1   $  \\[1mm]
$  350 $ & $  151.8   $ &  $  104.6   $ &  $   74.4   $ & $  215.2   $ &  $  181.1   $ &  $  148.4   $ & $  211.0   $ &  $  204.9   $ &  $  188.2   $  \\[1mm]
$  400 $ & $   37.9   $ &  $   25.8   $ &  $   18.2   $ & $   56.3   $ &  $   46.6   $ &  $   37.7   $ & $   56.2   $ &  $   53.9   $ &  $   48.9   $  \\[1mm]
$  450 $ & $    9.2   $ &  $    6.2   $ &  $    4.3   $ & $   14.3   $ &  $   11.6   $ &  $    9.2   $ & $   14.5   $ &  $   13.7   $ &  $   12.3   $  \\[1mm]
$  500 $ & $    2.1   $ &  $    1.4   $ &  $    1.0   $ & $    3.4   $ &  $    2.7   $ &  $    2.1   $ & $    3.6   $ &  $    3.3   $ &  $    2.9   $  \\[1mm]
\bottomrule
 \end{tabular}
    \caption{\small
      \label{tab:xsectevavalues}
      Same as Tab.~\ref{tab:xsecvalues} for Tevatron at 
      $\sqrt{s}=1.96\TeV$.}
\end{table}
%

%
% ---------------------------------------------------------------------------
%
\section{Conclusion and Summary}

In this letter we have investigated the effect of higher order soft corrections on
the total cross section for hadronic squark pair-production.
These radiative corrections make up numerically for a large part 
of the higher order QCD effects. 
Starting from the existing NLO calculation 
we provide for the ease-of-use the NLO scaling functions $f^{(10)}_{ij}$ in
the form of parameterizations which are accurate at the per mille level.
Subsequently we have employed well established techniques for soft gluon resummation to
derive new approximate NNLO results.
Our two-loop expressions for the scaling functions 
$f^{(20)}_{ij}$ are exact in all logarithmically enhanced terms near threshold 
and they include the Coulomb corrections. 
All two-loop scaling functions $f^{(21)}_{ij}$ and $f^{(22)}_{ij}$ governing 
the scale dependence have been computed exactly using renormalization group arguments.

For Tevatron and LHC, our approximate NNLO cross section leads 
to a cross section increase of $9\%$ compared to the NLO predictions, 
which translates into higher exclusion limits for squark masses.
Moreover, with our approximate NNLO result we have found significantly 
improved stability with respect to variation of the renormalization and 
factorization scale (keeping $\mur = \muf$). 
This leads to a residual theoretical uncertainty of $3\%$ plus a 
(largely uncorrelated) error due to the parton luminosity depending 
on the particular PDF set.

As a possible extension as far as the study of QCD corrections is concerned
we would like to mention bound-state effects for squark pair-production at hadron colliders,
which depend on the particular color and angular momentum quantum numbers
of the $\sq \sqb$-pair.
This would allow the resummation of the Coulomb corrections to all orders.
From similar recent work for top-quark pairs~\cite{Hagiwara:2008df,Kiyo:2008bv} 
we would expect a shift of the total cross section by ${\cal O}(1\%)$ due 
to these bound state corrections.

\section*{Acknowledgments}
We would like to thank T.~Plehn for discussions and help with 
the {\sc{Prospino}} code~\cite{Beenakker:1996ed}.
The Feynman diagrams have been prepared with {\sc{Axodraw}} \cite{Vermaseren:1994je} 
and for numerical integrations the {\sc{Cuba}}-library~\cite{Hahn:2004fe} has been used.
We acknowledge support by the Helmholtz Gemeinschaft under contract 
VH-NG-105 and in part by the Deutsche Forschungsgemeinschaft in 
Sonderforschungs\-be\-reich/Transregio~9.

%
% ---------------------------------------------------------------------
%
\renewcommand{\theequation}{A.\arabic{equation}}
\setcounter{equation}{0}
\section*{Appendix A: Useful formulae}

For the partonic cross section one can distinguish the following subprocesses for squark pair-production
according to the initial flavors and chiralities 
(see also Refs.~\cite{Bornhauser:2007bf,Gehrmann:2004xu,Bozzi:2005sy}):
\begin{eqnarray}
  \label{eq:uuuRuR}
  q \bar{q} &\to \, \sq_R \sqb_R,&\quad q{\phantom{,Q}} = u,d,s,c,b,\\
  \label{eq:uuuRuL}
  q \bar{q} &\to \, \sq_R \sqb_L,&\quad q{\phantom{,Q}} = u,d,s,c,b,\\
  \label{eq:uudRdR}
  q \bar{q} &\to \, \sQ_R \sQb_R,&\quad q, Q = u,d,s,c,b,\quad q \neq Q,\\
  \label{eq:uduRdR}
  q \bar{Q} &\to \, \sq_R \sQb_R,&\quad q, Q = u,d,s,c,b,\quad q \neq Q,\\
  \label{eq:uduRdL}
  q \bar{Q} &\to \, \sq_R \sQb_L,&\quad q, Q = u,d,s,c,b,\quad q \neq Q.
\end{eqnarray}
The same processes are possible if one replaces the chirality index $R$ with $L$ and vice versa.
The process~(\ref{eq:uuuRuR}) proceeds via gluon exchange in the $s$-channel and gluino exchange in the $t$-channel,
the process~(\ref{eq:uudRdR}) only via gluon exchange, and the
processes~(\ref{eq:uuuRuL}), (\ref{eq:uduRdR}), 
and (\ref{eq:uduRdL}) only via gluino exchange. 
The corresponding Born cross sections are given by
\begin{eqnarray}
%1
\hat{\sigma}(q \bar{q} \to \sq_R \sqb_R) & = &
      \frac{2 \* \pi \* \alpha^2_s}{27 \* \hats} \* \beta^3
      - \frac{2 \* \pi \* \yuk^2}{27 \* \hats} \*
            \left[6\*\beta +  
            3\*\left(1 + \frac{2\*(\mg^2-\mqR^2)}{\hats}\right)\*L_1\right]\notag\\[2mm]
          &&  +  \frac{2 \* \pi \* \as\yuk}{27 \* \hats} \*\left[
            \beta \* \left(1 + \frac{2\*(\mg^2-\mqR^2)}{\hats} \right)
             +2 \*\frac{(\mg^2-\mqR^2)^2 + \mg^2 \* \hats}{\hats^2}\*L_1\right]
          \, ,
\\[4mm]
%2
\hat{\sigma}(q \bar{q} \to \sq_R \sqb_L) & = &
     \frac{2\*\pi \*\yuk^2}{9} \* \frac{\mg^2}{\hats}\*
     \frac{\sqrt{(\mqR^2 + \mqL^2-\hats)^2- 4\*\mqR^2 \*\mqL^2}}
            {(\mg^2-\mqL^2)(\mg^2-\mqR^2)+\mg^2 \*\hats} 
          \, ,
\\[4mm]
%3
\hat{\sigma}(q \bar{q} \to \sQ_R \sQb_R) & = &\frac{2\*\pi \*\alpha^2_s}{27 \*\hats}\*\beta^3
          \, ,
\\[4mm]
%4
\hat{\sigma}(q \bar{Q} \to \sq_R \sQb_R) & = &
      -\frac{2\*\pi \*\yuk^2}{9 \* \hats}\*\left[
       2\*\frac{\sqrt{(\mqR^2 + \mQR^2 - \hats)^2- 4\*\mqR^2 \*\mQR^2}}{\hats}\right.\notag\\[2mm]
      &&\qquad\qquad\qquad\left.
      +\left(1 + \frac{2\*\mg^2 - \mqR^2 -\mQR^2}{\hats}\right)\*L_1^\prime\right]
          \, ,
\\[4mm]
\hat{\sigma}(q \bar{Q} \to \sq_R \sQb_L) & = &
      \frac{2\*\pi\* \yuk^2}{9}\* \frac{\mg^2}{\hats}
      \frac{\sqrt{(\mqR^2 + \mQL^2 - \hats)^2- 4\*\mqR^2 \*\mQL^2}}
            {(\mg^2-\mQL^2)(\mg^2-\mqR^2)+\mg^2 \*\hats}
\end{eqnarray}
with $\mq$, $m_{\tilde{q}_{R/L}}$, and $m_{\tilde{Q}_{R/L}}$ denoting squark
masses, $L_1$ being defined in Eq.~(\ref{eq:L1def}), and $L_1^\prime$ given by
\begin{eqnarray}
L_1^\prime &=& \log\left(\dfrac{\hats + 2\mg^2 -\mQR^2 - \mqR^2 
                             -\sqrt{(\mqR^2 + \mQR^2-\hats)^2- 4\*\mqR^2 \*\mQR^2}}
                          {\hats + 2\mg^2 -\mQR^2 - \mqR^2 
                             + \sqrt{(\mqR^2 + \mQR^2-\hats)^2- 4\*\mqR^2 \*\mQR^2}}
    \right) .
\end{eqnarray}

\bigskip

We have parameterized the numerical result of {\sc{Prospino}}~\cite{Beenakker:1996ed} 
for the NLO scaling functions in $f^{(10)}_{gg}$, $f^{(10)}_{q\barq}$, and 
$f^{(10)}_{q_i \bar{q}_j}$ by a fit (accurate at the per mille level), 
where the fit function $h(\beta)$ in Eqs.~(\ref{eq:fit10}) and (\ref{eq:fitqq10}) reads
\begin{eqnarray}
\label{eq:hfitdef}
h(\beta)&=& \phantom{+}\beta\*\Big[ a_1 
            + a_2\*\beta 
            + a_3\*\beta^2\*\log(8\*\beta^2) 
            + a_4\beta^3\*\log(8\*\beta^2) \notag\\[2mm]
        & & + a_{5} \* \frac{1}{\sqrt{1+\eta}}\*\log^2(1+\eta)
            + a_{6} \*\frac{1}{\sqrt{1+\eta}}\*\log(1+\eta) 
            + a_{7} \* \frac{1}{1+\eta}\* \log\left(\frac{1+\beta}{1-\beta}\right) \notag\\[2mm]
        & & + a_{8} \*z\* \log^2(\eta) 
            + a_{9} \* z\* \log(\eta) 
            + a_{10} \*z^2\* \log^2(\eta) 
            + a_{11} \*z^2\* \log(\eta) \notag\\[2mm]
        & & + a_{12} \*z^3\* \log^2(\eta) 
            + a_{13} \*z^3\* \log(\eta) 
            + a_{14} \*z^4\* \log^2(\eta) 
            + a_{15} \*z^4\* \log(\eta) \notag\\[2mm]
        & & + a_{16} \*z^5\* \log^2(\eta)
            + a_{17} \*z^5\* \log(\eta) \Big]
            \, ,
\end{eqnarray}
and $z$ is defined as 
\begin{eqnarray}
  \label{eq:xfitdef}  
  z \,=\, \frac{\eta}{(1+\eta)^2} \, , \qquad\qquad
  \beta \,=\, \sqrt{\frac{\eta}{1+\eta}}
  \, .
\end{eqnarray}
We present the coefficients of $h(\beta)$ in Tab.~\ref{tab:fitcoeff}.
For our choice of squark and gluino masses, $\mq = 400\GeV$ and $\mg = 500\GeV$, 
the fit is accurate at the per mille level.
However, the squark mass dependence of the coefficients $a_i$ is weak, 
so that for the ratio $\mg = 1.25\mq$, 
the fit may even be used in the whole range $\mq = 200\GeV \dots 1\TeV$ with a 
an error of less than $1\%$, the maximum error being $0.7\%$ at the lowest values of $\mq$.
\begin{table}[ht!]
\centering
\begin{tabular}{lrrr}
\toprule
$a_i$ & $f_{gg}^{(10)}\phantom{000}$& $f_{q\bar{q}}^{(10)}\phantom{000}$&
$f_{q_{i}\bar{q}_{j}}^{(10)}\phantom{000}$ \\[1mm]
\hline
$a_{ 1 }$& $   0.49100112  $ & $   0.14999587 $ & $   0.14422667 $  \\[1mm]
$a_{ 2 }$& $  -1.91412644  $ & $   7.76393459 $ & $   1.63082127 $  \\[1mm]
$a_{ 3 }$& $  -0.60192840  $ & $   0.67354296 $ & $  -0.23767546 $  \\[1mm]
$a_{ 4 }$& $   1.55221239  $ & $  -4.45797953 $ & $  -0.59322814 $  \\[1mm]
$a_{ 5 }$& $  -0.00562902  $ & $   0.01846674 $ & $  -0.00167924 $  \\[1mm]
$a_{ 6 }$& $   0.05091553  $ & $  -0.16690138 $ & $   0.01653721 $  \\[1mm]
$a_{ 7 }$& $   0.96279970  $ & $  -4.26865285 $ & $  -0.88130118 $  \\[1mm]
$a_{ 8 }$& $  -0.32470009  $ & $   0.61779585 $ & $   0.07212666 $  \\[1mm]
$a_{ 9 }$& $  -0.12126013  $ & $   1.09469670 $ & $   0.36722998 $  \\[1mm]
$a_{ 10 }$& $  -0.50416435  $ & $   0.85492550 $ & $   0.03809051 $  \\[1mm]
$a_{ 11 }$& $   0.70967285  $ & $  -0.61743437 $ & $  -0.48810751 $  \\[1mm]
$a_{ 12 }$& $   3.99890940  $ & $  -8.15404313 $ & $  -0.02789299 $  \\[1mm]
$a_{ 13 }$& $  -9.83101731  $ & $   4.99311643 $ & $   2.37790285 $  \\[1mm]
$a_{ 14 }$& $ -16.88306230  $ & $  44.33618984 $ & $  -0.60307995 $  \\[1mm]
$a_{ 15 }$& $  52.83111706  $ & $ -26.25150811 $ & $  -7.30816582 $  \\[1mm]
$a_{ 16 }$& $  20.74447249  $ & $ -93.67735527 $ & $   1.80344717 $  \\[1mm]
$a_{ 17 }$& $ -99.69494837  $ & $  57.74025069 $ & $   9.30489420 $  \\[1mm]
\bottomrule
\end{tabular}
\caption{\small
  \label{tab:fitcoeff}
  Coefficients for the fit functions~(\ref{eq:fit10}), (\ref{eq:fitqq10}) and 
  (\ref{eq:hfitdef}) for $\mq = 400\GeV$ and $\mg = 500\GeV$.
  }
\end{table}

Note that the values for the constant $a_1$ for quark-anti-quark scattering 
in Tab.~\ref{tab:fitcoeff} can also be used for other ratios of 
$r_q = \mg/\mq$ to very good accuracy because of the weak squark mass dependence at NLO.
Since $a_1 = a_1(r_q)$, we can be extract from Tab.~\ref{tab:fitcoeff} 
for the value of $a_1$ at $r_q = 1.25$:
\begin{eqnarray}
\label{eq:a1rqdef}
a_1(r_q) = \frac{1681}{400}\frac{r_q^2}{(1+r_q^2)^2}\,\, a_1(r_q=1.25)\, .
\end{eqnarray}
This is due to $\lim_{\beta \to 0} h(\beta)/\beta = a_1$ and Eq.~(\ref{eq:fqq00}).

%
% ---------------------------------------------------------------------
%
{\footnotesize
% \bibliographystyle{h-elsevier2}
% \bibliography{litsqprod}

}

\end{document}